\documentclass[journal]{vgtc}                     

\onlineid{1177}

\vgtccategory{Research}

\title{VisPuzzle: Task-Aware Composite Visualization Construction}

\author{%
  \authororcid{Zheng Wang*}{0009-0004-0575-5936},
  \authororcid{Zhiyang Shen*}{0009-0006-7644-5083},
  \authororcid{Lingyun Yu}{0000-0002-3152-2587
}, and
  \authororcid{Shixia Liu}{0000-0003-4499-6420}
}

\authorfooter{
  \item
    Z.~Wang, Z.~Shen, and S.~Liu are with the School of Software, BNRist, Tsinghua University. Z.~Wang and Z.~Shen are joint first authors. S.~Liu is the corresponding author.
    E-mail: \{\{wangz24, shenzhiy21\}@mails., shixia@\}tsinghua.edu.cn.
    \item L.~Yu is with Xi'an Jiaotong-Liverpool University.
    E-mail: Lingyun.Yu@xjtlu.edu.cn.
}

\abstract{Compositing multiple visualizations into a coherent whole remains challenging due to the vast design space and the need to balance the coverage of task-relevant data insights (\eg trends and outliers), perceptual clarity, and aesthetic quality.
In this paper, we present \sys, a task-aware method that formulates visualization composition as a stepwise search problem over a composition graph.
In this graph, nodes represent either data composition operations (\eg union, join) or visual composition operations that determine component relationships, spatial arrangements, or component proportions, and edges encode feasible transitions between operations.
We employ Monte Carlo Graph Search to efficiently identify high-quality composition candidates from this graph, guided by a reward function that balances 
task relevance, perceptual effectiveness, and aesthetic coherence. 
A use case and a user study show that the top-ranked candidates produced by \sys align closely with human judgments of composition quality, demonstrating its utility in supporting principled and scalable visualization composition. 

\vspace{1mm}
\noindent\textbf{Code}: \url{https://github.com/vispuzzle/vispuzzle}. 
\hspace{4mm}
\noindent\textbf{Corpus}: \url{https://huggingface.co/datasets/vispuzzle/vispuzzle}.
}


\teaser{
  \centering
  \includegraphics[width=\linewidth]{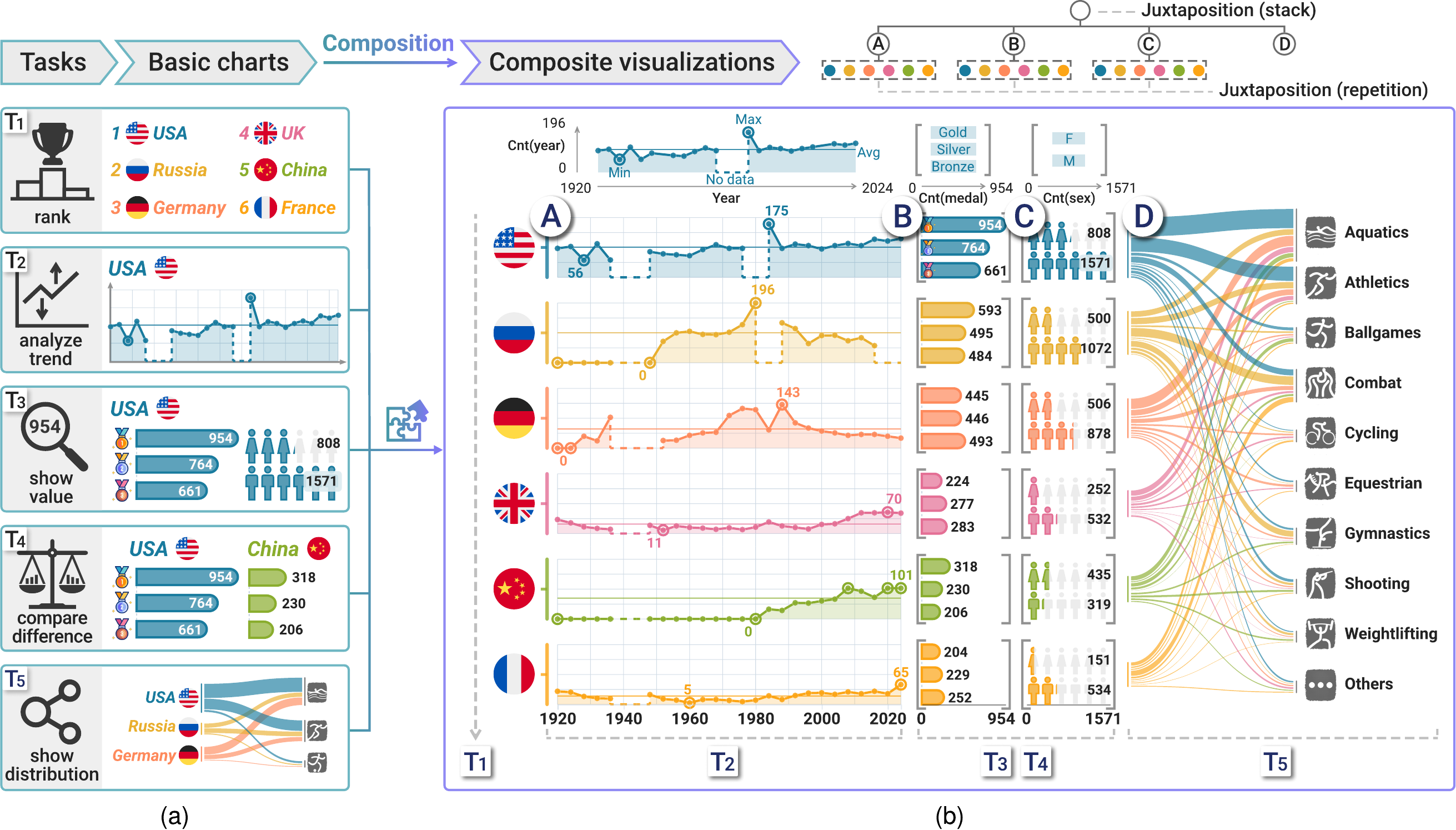}
  \caption{Task-driven composite visualization construction: given analysis tasks, \sys (a) selects appropriate basic charts to reveal relevant data insights, and (b) integrates them to construct a composite visualization design, balancing task relevance, perceptual effectiveness, and aesthetic coherence.
  The resulting design vertically organizes six top medal-winning countries, and horizontally aligns line charts, bar charts, and an alluvial diagram. It presents multifaceted insights into the Summer Olympic Medals dataset, including medal trends, gender differences, and sporting strengths across countries.
  }
  \label{fig:teaser}
}

\graphicspath{{figs/}{figures/}{pictures/}{images/}{./}} 

\usepackage{enumerate}
\usepackage{color}
\usepackage{lipsum}                    
\usepackage{amssymb,amsmath}
\usepackage{wrapfig}
\usepackage{booktabs}
\usepackage{makecell}
\usepackage{bm}
\usepackage{newtxtext}
\usepackage{array}
\usepackage{xspace}
\usepackage{xcolor}
\usepackage{url}
\usepackage{multirow}
\usepackage{graphicx}
\usepackage{stfloats}
\usepackage{svg}
\usepackage{tabularx}
\usepackage[normalem]{ulem}
\usepackage{xurl}
\usepackage{tcolorbox}
\definecolor{boxbg}{RGB}{246, 248, 255}

\usepackage{algorithm}
\usepackage{algpseudocode}

\def \etal {{\emph{et al}.\thinspace}}
\def \eg {{\emph{e.g.},\thinspace}}
\def \ie {{\emph{i.e.},\thinspace}}
\definecolor{darkgreen}{rgb}{0,0.5,0}
\newcommand{\stab}{\vspace{1.2ex}\noindent}
\newcommand{\stitle}[1]{\stab\noindent{\textbf{#1}}}

\newcommand{\vs}{\textit{vs.}\xspace}

\newcommand{\sys}{\textbf{\textsc{VisPuzzle}}\xspace}
\newcommand{\numDataset}{{2,055}\xspace}

\newcommand{\imgtext}[3][0.9cm]{
    \noindent
    \begin{minipage}[t]{0.1\columnwidth}
        \vspace{0pt}
        \centering
        \includegraphics[width=#1]{#2} 
    \end{minipage}
    \hfill
    \begin{minipage}[t]{0.86\columnwidth}
        \vspace{0pt}
        #3
    \end{minipage}
    \par             
    \vspace{-\parskip}
    \vspace{5pt}
}

\usepackage{mathptmx}                  
\DeclareMathAlphabet{\mathcal}{OMS}{cmsy}{m}{n}
\DeclareSymbolFont{largesymbols}{OMX}{cmex}{m}{n}

\begin{document}


\fontsize{9}{9} 
\firstsection{Introduction}
\maketitle

As datasets grow in scale and heterogeneity, analysts increasingly need to examine multiple complementary facets, such as temporal trends, categorical breakdowns, and subgroup comparisons, to gain a comprehensive understanding. 
Although visualization is effective for presenting information, a single chart is often insufficient to convey these facets simultaneously~\cite{munzner2014visualization}.
To address this limitation, visualization composition integrates multiple visualization components into a coherent, aesthetically pleasing layout~\cite{deng2023revisiting, javed2012exploring}, enabling analysts to inspect multiple aspects in a unified view.
For example, Fig.~\ref{fig:teaser}(b) shows such a design for the Summer Olympic Medals dataset, where coordinated components, including line charts, bar charts, and an alluvial diagram, jointly support both within-country analysis (\eg long-term performance evolution and gender distribution) and cross-country comparison (\eg performance differences and sport-specific strengths).

Despite their analytical promise, designing effective composite visualizations is challenging. 
It requires balancing three interrelated criteria~\cite{kirk2016data}:
1) \textbf{task relevance}, the coverage of the data insights required by the analysis tasks; 2) \textbf{perceptual effectiveness}, the use of perceptually clear layouts to support accurate interpretation with minimal cognitive effort; and 3) \textbf{aesthetic coherence}, the visual harmony and stylistic consistency among constituent components.
These criteria are complementary but often competing.
For example, increasing task relevance by incorporating additional task-relevant insights (\eg trends and outliers) can easily lead to visual clutter, thereby compromising perceptual clarity.
Conversely, layouts optimized for clarity may limit the diversity of insights, while enforcing aesthetic coherence can constrain encoding and layout flexibility, limiting the capacity to represent certain insights.\looseness=-1

This challenge stems from the need to coordinate decisions at both the data and visual levels. 
Traditionally, at the data level, designers first decompose the input into meaningful subsets to reveal data insights, then combine them based on their relationships.
At the visual level, they determine appropriate visual mappings, composition structures, layouts, and styles (\eg color, typography, rendering) to present these insights clearly and coherently.
These interdependent decisions create a large, complex design space, making trade-offs difficult to manage.
As a result, achieving a proper balance among task relevance, perceptual effectiveness, and aesthetic coherence remains difficult.
While manual design can produce high-quality results~\cite{lex2014upset, henry2007nodetrix, collins2007vislink}, it relies heavily on expertise and iterative refinement, making the process time-consuming and difficult to scale~\cite{deng2023revisiting}.
To the best of our knowledge, there is no automatic method that can effectively manage these trade-offs to construct composite visualizations of comparable quality.

To fill this gap, we present \sys, a task-aware visualization composition method that jointly optimizes data composition and visual composition.
To guide the automatic construction process, we collect \numDataset composite visualization designs from three sources: 1) a curated composite visualization dataset~\cite{deng2023revisiting}, 2) complementary research publications, and 3) public design sources (\eg Visual Capitalist~\cite{visualcapitalist} and Pinterest~\cite{pinterest}).
Through a comprehensive analysis of these visualizations, we extend the design space developed by Deng~\etal~\cite{deng2023revisiting} to characterize composition along three dimensions: component relationship, spatial arrangement, and component proportion.
Based on the design space, we formulate the composition process as a stepwise search problem over a composition graph.
In this graph, nodes represent either data composition operations (\eg union, join) or visual composition operations that determine component relationships, spatial arrangements, or component proportions, and edges encode feasible transitions between operations.
We employ Monte Carlo Graph Search (MCGS) to efficiently identify high-quality composition candidates, guided by a reward function balancing task relevance, perceptual effectiveness, and aesthetic coherence. 
We showcase the capabilities of \sys through a use case and evaluate composition quality via a user study.
The study results show that the top-ranked candidates align well with human judgments of composition quality.

In summary, the primary contributions of this work are:
\begin{itemize}[nosep]
    \item A corpus of \numDataset representative composite visualization designs. 
    \item An extended design space that characterizes the step-by-step construction of composite visualizations.
    \item A task-aware method for composite visualization construction via joint optimization of task relevance, perceptual effectiveness, and aesthetic coherence.
\end{itemize}

\section{Related Work}
\label{sec:related}
Given the demonstrated value of composite visualizations, practitioners have developed many manual designs.
As early as 1801, William Playfair's seminal work demonstrated the utility of this paradigm by combining charts, such as pie and bar charts, to effectively communicate multifaceted economic data~\cite{playfair1801statistical}.
This composition strategy has continued to influence current visualization designs.
For example, UpSet~\cite{lex2014upset} aligns a matrix visualization with bar charts and box plots along a shared axis, which facilitates the exploration of complex set relationships. 
Similarly, VisLink~\cite{collins2007vislink} connects multiple components with explicit links to reveal their data relationships, while TextFlow~\cite{cui2011textflow} overlays visual components onto a topic flow graph to display data in context.
Other designs, such as NodeTrix~\cite{henry2007nodetrix}, MizBee~\cite{meyer2009mizbee}, CNNVis~\cite{liu2017towards}, DataLinker~\cite{chen2021interactive}, FSLDiagnotor~\cite{yang2022diagnosing}, and Aardvark~\cite{lange2025aardvark}, embed specific visual components (\eg matrices, scatter plots) within the visual elements of other visualizations (\eg node-link diagrams, voronoi diagrams) to balance overview and detail within a single composite view.
\looseness=-1

Although these visualizations have demonstrated potential in many applications, their design remains an expertise-intensive process~\cite{chen2020composition, gleicher2011visual, munzner2014visualization, wang2018visualization}. 
Designers need to coordinate multiple interdependent decisions, including data selection, visual mappings, composition structures and layouts, and styling choices, so that different components collectively support coherent analysis.
To facilitate this process, efforts have been made to systematize the design space for composite visualizations, enabling structured and reproducible design~\cite{javed2012exploring, deng2023revisiting, ying2024vaid, zhu2024compositing, shi2026piccl}. 
For example, Javed and Elmqvist~\cite{javed2012exploring} characterized composite visualizations by the spatial and data relationships among visual components and identified five common component relationships: juxtaposition, integration, overloading, superimposition, and nesting. 
Building on this work, Deng~\etal~\cite{deng2023revisiting} analyzed composite visualizations collected from IEEE VIS publications between 2006 and 2020 and refined the taxonomy. 
They summarized eight component relationships: repetition, mirror, stack, co-axis, coordinate, annotation, large panel, and nesting, and organized them under three strategies: juxtaposition, overlay, and nesting. 
Recently, Zhu~\etal~\cite{zhu2024compositing} extended this line of work to facilitate the design of composite visualizations in immersive environments through embodied interactions, such as grabbing and colliding.

While these studies have systematized the component relationships for composite visualizations, they do not explicitly model the geometric aspects for the step-by-step composition process, including spatial arrangement and the proportion of space allocated to components, which are critical for developing an automatic composition method.
As a result, designing such visualizations still relies heavily on manual effort to assemble visual components for specific analysis requirements.
To address this issue, we first extend the design space from Deng~\etal~\cite{deng2023revisiting} to explicitly model the step-by-step composition process.
Then, we formulate the composition process as a stepwise search problem and employ MCGS to automatically construct high-quality compositions.

\section{Design Goals}
\label{sec:designgoals}
The design of \sys is informed by both visualization design literature and expert practice. 
Drawing on prior visualization research on task abstraction~\cite{brehmer2013multi, munzner2009nested}, layout organization~\cite{lu2020exploring, wang2000guidelines}, and aesthetics~\cite{harrison2015infographic}, we identify three key considerations around composite visualization design: what task-relevant insights to include, how to ensure a perceptually effective layout, and how to maintain aesthetic coherence across multiple components. 
These considerations correspond to three criteria: \textbf{task relevance}, \textbf{perceptual effectiveness}, and \textbf{aesthetic coherence}.
To ground these literature-derived criteria with practical design experience, we interviewed four experts ($E_1$--$E_4$), none of whom are co-authors of this work. 
We selected the experts based on two criteria: expertise in visualization design or automatic visualization and years of relevant experience.
$E_1$ and $E_2$ are Ph.D. students with five and eight years of experience in visualization design, respectively, and have manually crafted composite visualizations for several visualization and machine learning projects. 
$E_3$ is a professor with ten years of experience in visual analytics and infographic generation, and $E_4$ is a professor with eight years of experience in data intelligence and automatic visualization. 
Each interview lasted 60--90 minutes and followed a semi-structured format, covering the experts' design considerations, authoring workflows, and practical challenges.
Two authors independently reviewed the interview notes using thematic analysis and resolved disagreements through discussion.
By synthesizing insights from the literature and interviews, we identified three design goals to guide the development of our method.
\looseness=-1

\textbf{G1: Ensure task relevance by revealing task-relevant data insights}. 
Task relevance in visualization is concerned with whether the visual representation conveys the data insights (\eg trends and outliers) most relevant to the user's analysis tasks (\eg characterizing medal trends for specific years)~\cite{munzner2009nested, brehmer2013multi}.
Achieving this in composite visualizations becomes more demanding, as multiple visual components must be coordinated to collectively communicate complementary insights rather than fragmented information. 
All experts emphasized that deciding which data aspects should be visualized and how insights should be distributed and coordinated across visual components is a primary design consideration.
For example, $E_3$ noted: ``When designing composite visualizations, I often need to carefully consider how to distribute insights across multiple charts so that they complement rather than duplicate each other.'' 
This finding motivates the need to support the selection and coordination of visual components to best reveal task-relevant insights.
\looseness=-1

\textbf{G2: Ensure a perceptually effective layout to support efficient perception and interpretation}.
Perceptual effectiveness in visualization refers to whether the layout of visual components supports accurate interpretation and efficient understanding~\cite{wang2000guidelines, lu2020exploring, yang2025dashboard}.
Achieving such perceptual effectiveness in composite visualizations becomes more complex because integrating multiple charts increases information density, introduces competing visual signals, and increases the risk of visual clutter.
Without careful layout organization, poorly structured layouts can obscure structural relationships among components and disrupt the clarity of information flow.
This issue was consistently raised by our experts. 
For example, $E_1$ commented: ``Without careful layout organization, visual clutter can quickly overwhelm users and make it difficult to maintain a coherent understanding of how information is conveyed across components.''
This observation highlights the need to generate well-structured layouts that support clear analytical narratives and efficient interpretation across multiple components.\looseness=-1

\textbf{G3: Maintain aesthetic coherence through visual harmony and stylistic consistency}.
Aesthetic coherence is the harmonious integration of visual elements into a unified and visually consistent design~\cite{harrison2015infographic}.
Prior research has shown that such coherence enhances user engagement and memorability~\cite{borkin2013makes}.
In composite visualizations, achieving aesthetic coherence requires integrating different visual encodings and stylistic conventions into a cohesive whole.
Our experts noted that achieving such coherence often demands careful aesthetic judgment and iterative refinement.
For example, $E_4$ pointed out: ``In composite visualizations, achieving a consistent visual style often requires deliberate adjustments. 
I frequently refine colors, spacing, and visual emphasis across charts to ensure they form a visually coherent whole.''
This finding suggests that our method should promote visual harmony and stylistic consistency across components to improve the overall user experience.

Achieving these goals requires modeling visualization composition as a step-by-step process that determines how components are related and arranged, and how visual space is allocated to support perceptual effectiveness and aesthetic coherence.
This motivates the development of a design space for visualization composition.
\section{Extending Existing Design Space}
\label{sec:designspace}
Deng~\etal~\cite{deng2023revisiting} introduced a two-level taxonomy comprising eight component relationships.
However, component relationships alone do not specify how components are spatially arranged or how visual space is allocated, which directly affect perceptual effectiveness (\textbf{G2}) and aesthetic coherence (\textbf{G3}).
We therefore extend this taxonomy by incorporating geometric aspects, including spatial arrangement and component proportion, to model the step-by-step composition process.

\begin{figure*}[!tb]
    \centering
    \includegraphics[width=\textwidth]{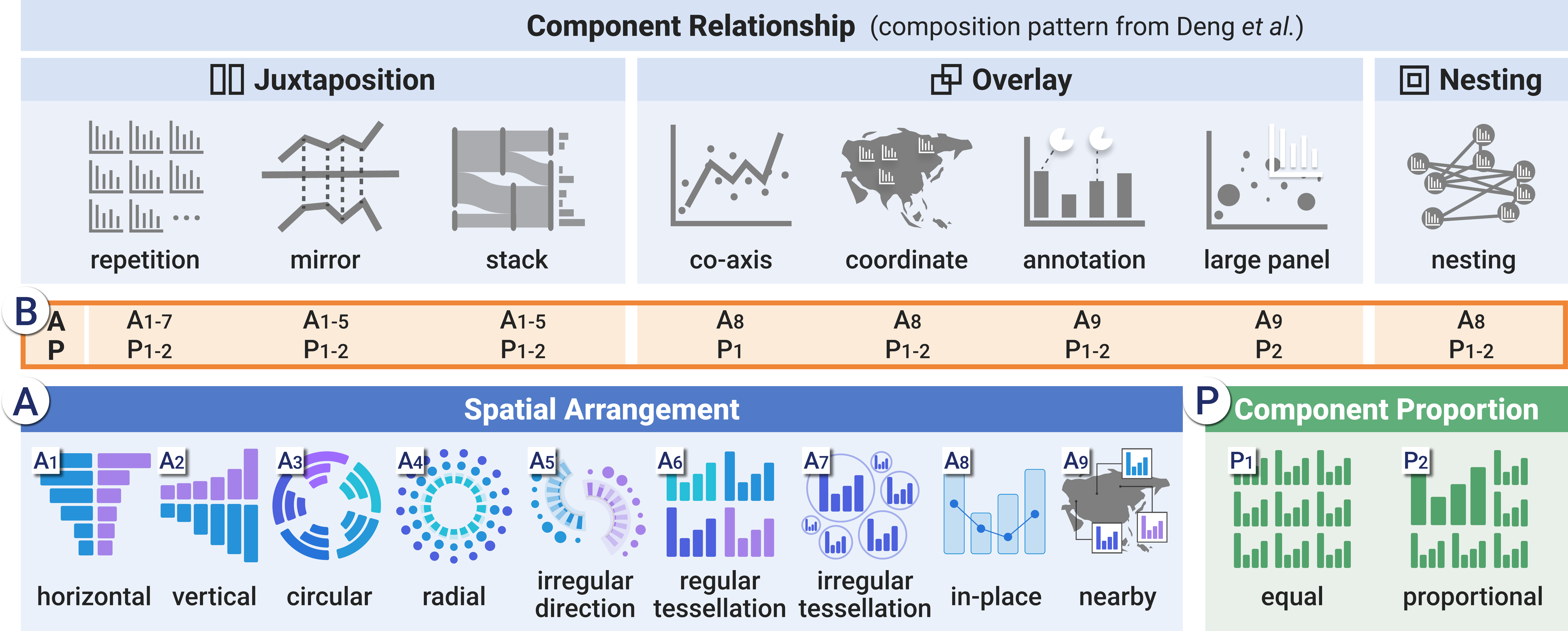}
    \caption{Extended composition design space: The upper row summarizes the \textit{component relationships} (composition patterns from Deng~\etal~\cite{deng2023revisiting}). The lower panels enumerate the set of \textit{spatial arrangement} (A) and \textit{component proportion} (P) schemes extended in our work. For each component relationship, the orange box (B) specifies its suitable \textit{spatial arrangements} and \textit{component proportions}, reflecting common compositional practice.}
    \label{fig:designspace}
    \vspace{-4mm}
\end{figure*}

\subsection{Data Collection}
\label{subsec:corpus}
The existing taxonomies are primarily derived from visualization research publications.
To capture a broad range of compositional practices, we constructed a corpus of \numDataset visualization designs, where each design contains at least one composite visualization. 
During data collection, we collaborated with experts $E_2$ and $E_3$ described in Sec.~\ref{sec:designgoals} to assess design quality, excluding low-resolution designs and designs without clear visualization compositions.
The corpus draws on three sources:
\looseness=-1

\begin{itemize}[nosep, wide=\parindent]
    \item \textit{A curated composite visualization dataset} ($N=866$) built by Deng~\etal~\cite{deng2023revisiting}, which consists of 866 visualization designs collected from IEEE VIS publications between 2006 and 2020. 
    This source provides continuity with the existing design space.
    
    \item \textit{Complementary research publications} ($N=500$). 
    To capture the recent developments in composite visualization design, we collected 500 examples from IEEE VIS publications between 2021 and 2025.

    \item \textit{Public design sources} ($N=689$). 
    To further diversify the design space, we additionally collected 689 infographic-style examples of composite visualizations from public online sources, such as \textit{Pinterest}~\cite{pinterest} and \textit{Visual Capitalist}~\cite{visualcapitalist}.
\end{itemize}

\subsection{Review and Analysis Process}
\label{subsec:review}
Following common practice~\cite{solen2026design, yao2025designing}, we analyzed the visualization designs through a three-phase process:

\stitle{Dimension initialization}.
As shown in Fig.~\ref{fig:designspace}, we initialized the design space based on the component relationships identified by Deng~\etal~\cite{deng2023revisiting}, \textit{juxtaposition}, \textit{overlay}, and \textit{nesting}, which are referred to as composition patterns in their work.
To examine how composite visualizations are constructed in practice, two authors independently analyzed a subset of the corpus with 450 visualization designs, 150 from each source.
Following the analysis process of Deng~\etal, we first decomposed each design into basic visualization components (\eg a bar chart).
We then annotated the spatial relationships among these components to characterize how they are combined. 
Through this process, we identified 682 composite visualizations from the subset of 450 visualization designs.
Building on these annotated compositions, the two authors independently investigated the layout strategies used to integrate visual components, as they directly reflect how components can be assembled during composition.
Through this analysis, the two authors identified two additional dimensions, \textit{spatial arrangement} and \textit{component proportion}, along with their preliminary categories.

\stitle{Dimension and category iteration}.
In the iteration phase, we refined these dimensions and their categories through multiple rounds of coding and comparison. 
In each round, the two authors independently applied the current design space to a new subset of the corpus comprising 300 visualization designs to assess: 1) whether additional layout strategies should be introduced as new dimensions, and 2) whether new categories emerged within the existing dimensions. 
When an example cannot be adequately characterized by the current design space, the authors discussed potential refinements or extensions to improve its coverage.
The design space was then updated accordingly and used in the next round, until no additional dimensions and categories were required.
\looseness=-1

\stitle{Categorization checking}.
After finalizing the dimensions and categories, each composition can be defined as an ordered tuple of the three dimensions: (\textit{component relationship}, \textit{spatial arrangement}, and \textit{component proportion}). 
To assess the coverage of the finalized design space, we applied the three dimensions and categories to the entire corpus of \numDataset composite visualization designs and obtained 3,149 tuples.
We verified whether all examples can be consistently described without introducing additional categories. 
Any remaining ambiguities in categorization were resolved through discussion among all authors.

\subsection{Extended Design Space}
\label{subsec:extendedspace}

As shown in Fig.~\ref{fig:designspace}, the resulting design space characterizes the construction of composite visualizations along three dimensions: component relationship, spatial arrangement, and component proportion.

\stitle{Component relationship}.
This dimension characterizes the spatial relationships among components.
It aligns with the design space proposed by Deng~\etal~\cite{deng2023revisiting}, which categorizes component relationships based on how visualizations spatially overlap, align, and relate to one another:
\looseness=-1

\begin{itemize}[nosep, wide=\parindent]
    \item \textit{Juxtaposition}, where components are placed side by side. This includes three sub-categories: \textit{Repetition} (identical visual structures; 38.2\%, 1,204 / 3,149), \textit{Mirror} (symmetrical layout; 2.8\%, 88), and \textit{Stack} (sequential placement along a shared axis or by identical data items; 21.9\%, 689).
    \item \textit{Overlay}, where components are placed on top of one another. This includes three sub-categories: \textit{Co-axis} (sharing a common coordinate system; 8.0\%, 251), \textit{Coordinate} (where one component provides the coordinate system for others; 9.3\%, 292), \textit{Annotation} (connecting child components to the visual elements of parent components; 2.3\%, 74), and \textit{Large panel} (directly overlaying child components on parent components without explicit visual links; 3.2\%, 102). 
    \item \textit{Nesting}, where child components are embedded into parent components' visual elements to form hierarchical structures (14.3\%, 449).\looseness=-1
\end{itemize}

\stitle{Spatial arrangement}.
This dimension describes how components are geometrically organized within the view.
We identify nine spatial arrangements in total:

\imgtext{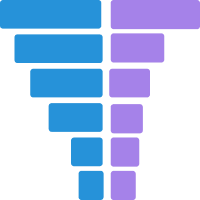} {
\textit{Horizontal} (22.5\%, 707). The \textit{horizontal} arrangement places components side by side. 
It is commonly used in Cartesian coordinate systems to present sequential visualizations along the horizontal axis.
}
\imgtext{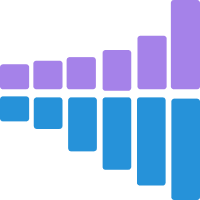} {
\textit{Vertical} (19.8\%, 623). The \textit{vertical} arrangement places components above one another. 
Similar to the horizontal arrangement, it is used in Cartesian coordinate systems to organize sequential components along the vertical axis.
}
\imgtext{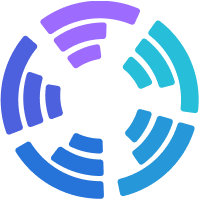} {
\textit{Circular} (1.6\%, 49). The \textit{circular} arrangement positions components along a circular path. 
It is commonly used in polar coordinate systems to organize components around a central reference.
}
\imgtext{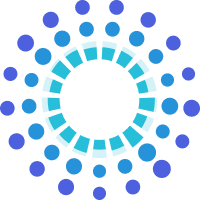} {
\textit{Radial} (3.7\%, 116). The \textit{radial} arrangement radiates components outward from a central reference. 
It often uses the center as a hub, guiding users' attention from a core concept outward to surrounding components.
}
\imgtext{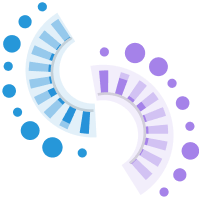} {
\textit{Irregular direction} (1.3\%, 40). The \textit{irregular direction} arrangement positions components along free-form paths without a fixed structural pattern. 
It is often used to create flexible, aesthetically pleasing layouts that prioritize visual appeal and creative freedom.
}
\imgtext{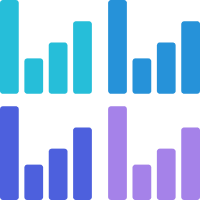} {
\textit{Regular tessellation} (11.5\%, 362). The \textit{regular tessellation} arrangement places components in a regular, repeating grid with equal spacing.
It ensures the uniform coverage of the display space, producing a balanced and ordered visual appearance.\looseness=-1
}
\imgtext{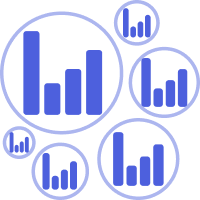} {
\textit{Irregular tessellation} (2.7\%, 84). The \textit{irregular tessellation} arrangement places components without a uniform grid, allowing flexible positioning based on relationships in the data. \looseness=-1
}
\imgtext{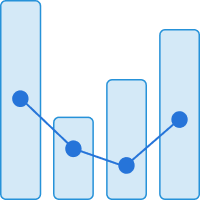} {
\textit{In-place} (31.5\%, 992). The \textit{in-place} arrangement places components within the same spatial region, often overlapping or co-located. 
It is commonly used in overlay and nesting relationships, where components share a common spatial reference to reveal relationships between them.
}
\imgtext{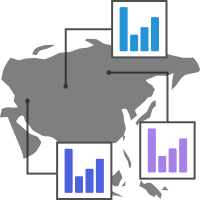} {
\textit{Nearby} (5.6\%, 176). The \textit{nearby} arrangement places secondary components close to their corresponding primary components without strict alignment. 
It is often used for annotations and large panels to associate related components.
}

\stitle{Component proportion}.
This dimension determines how much space is allocated to each component. Allocating more space to important components makes them more noticeable and attracts greater user attention.
We identify two proportion categories:

\imgtext{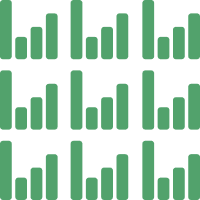} {
\textit{Equal} (68.6\%, 2,160). Components are uniformly distributed and allocated equal display space. It is typically used when components have equal importance or to support unbiased comparisons (\eg small multiples), ensuring that spatial size does not influence the perception of data magnitude.
}
\imgtext{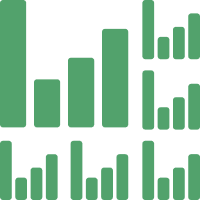} {
\textit{Proportional} (31.4\%, 989). Display space is allocated to components according to their relative importance. It is often used to guide user attention, with important components occupying more space to present detailed information, while less important components provide supporting or contextual content.\looseness=-1
}

Fig.~\ref{fig:designspace}B summarizes the spatial arrangements and component proportions suitable for each component relationship, which are obtained by a comprehensive analysis of the collected composite visualizations. 
Visual examples and their composition configurations along these three dimensions are provided in Supp.~A.

\section{Task-Aware Composition Method}
\label{subsec:selection}

\begin{figure*}[!t]
    \centering
    \includegraphics[width=\textwidth]{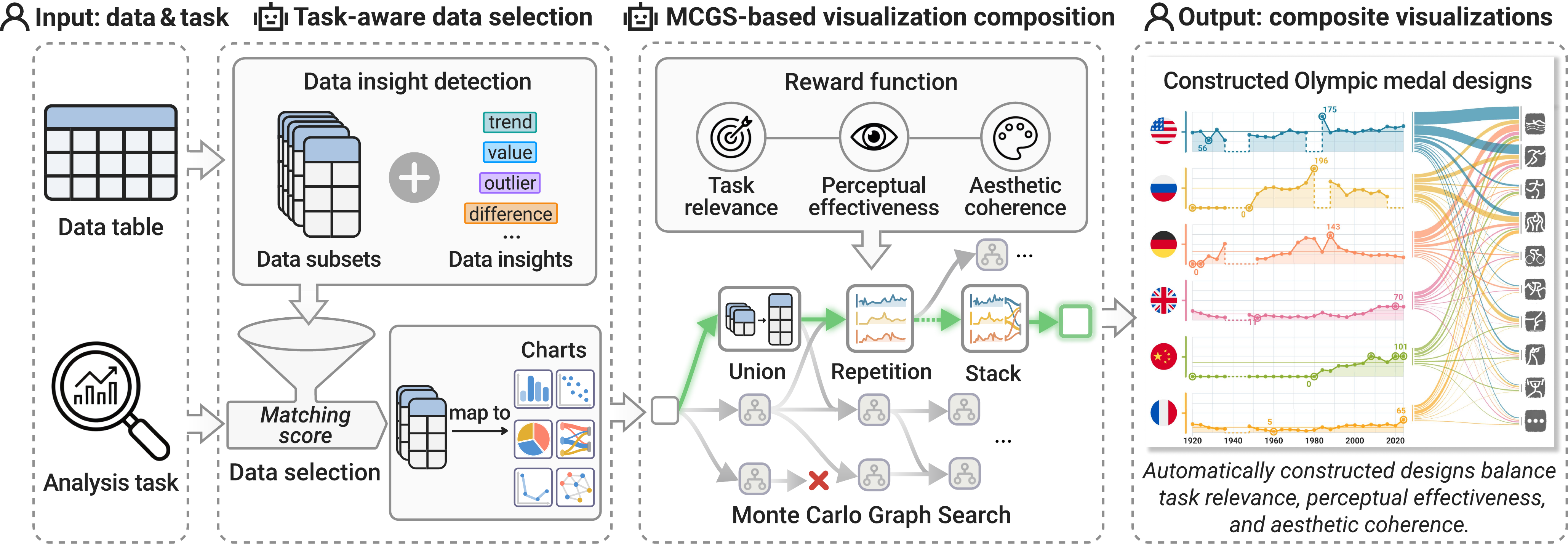}
    \caption{\sys Overview. Given an input data table and an analysis task, our method decomposes the data table into subsets, selects those that best reveal task-relevant insights, and maps them to appropriate basic charts. Building on the selected subsets and their charts, it employs the MCGS algorithm to efficiently identify high-quality composition candidates, guided by a reward function that jointly optimizes task relevance, perceptual effectiveness, and aesthetic coherence.
    Users can then review these candidates and select a preferred design. \smash{\includegraphics[height=2ex]{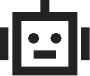}} indicates automated steps, and \smash{\includegraphics[height=2ex]{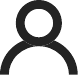}} indicates steps involving user interaction.}
    \label{fig:pipeline}
    \vspace{-4mm}
\end{figure*}

Guided by the design goals and the extended design space, we develop a task-aware visualization composition method that balances three criteria: task relevance (\textbf{G1}), perceptual effectiveness (\textbf{G2}), and aesthetic coherence (\textbf{G3}).
As shown in Fig.~\ref{fig:pipeline}, given an analysis task that specifies target data insights, our composition method automatically performs the following two steps. 
First, it selects subsets from the input data table to reveal task-relevant insights (\textbf{G1}), and then maps them to basic charts.
Second, building on these subsets and charts, we formulate the composition process as a stepwise search problem over a composition graph derived from the extended design space and solve it using Monte Carlo Graph Search (MCGS).
The search is guided by a reward function that jointly optimizes the three criteria.
After candidate composite visualization designs are constructed, users can either select a preferred design or refine the task for another round of construction.

\subsection{Task-Aware Data Selection}
\label{sec:selection}

Data selection begins by decomposing the input data table into candidate subsets formed by combinations of task-relevant data attributes. 
These subsets are further enriched through feasible transformation operations (\eg aggregation, binning, grouping) to generate additional variations.
From this enriched set, we extract task-relevant insights from each subset and select those that best reveal them.

\stitle{Data insight detection}. 
Following prior work~\cite{shi2021calliope, munzner2014visualization}, \sys supports ten common types of insights: value, difference, proportion, trend, categorization, distribution, rank, association, extreme, and outlier. 
We identify these insights using hypothesis testing, which detects statistically significant patterns within each subset~\cite{wang2020datashot}.
For example, an \textit{outlier} insight is detected by identifying data items that deviate significantly from the overall distribution.
To support downstream subset selection, each insight $I$ is complemented with two properties: \textit{statistical strength} and \textit{text description}.
Statistical strength $s_{\text{stat}}(I)$ is a value between 0 and 1 that quantifies how strongly the detected insight is supported by the data. 
It is computed using different statistical measures depending on the insight type, following Li~\etal~\cite{li2026chartgalaxy}.
For example, \textit{distribution} insights use goodness-of-fit scores to measure how well the data matches a candidate statistical distribution, and \textit{difference} insights use effect-size metrics to quantify the magnitude of differences between two groups of data.
Text description is generated using predefined templates to summarize the insight in natural language, following Shi~\etal~\cite{shi2021calliope}.
For example, a \textit{trend} insight can be described as ``\textit{medal count decreases first and then increases over time.}''
Further details about these two properties are provided in Supp.~B.

\stitle{Subset selection}.
To guide subset selection, we calculate a \textit{matching score} to quantify the alignment between a target insight $T$ from the task and a subset $S$ with detected insights $\mathcal{I}_S$:

\vspace{-2mm}
\begin{equation}
M(T,S) = \mathbb{I}_{\text{attr}}(T,S) \cdot \max_{I\in\mathcal{I}_S}(s_{\text{stat}}(I) \cdot s_{\text{cont}}(T, I)).
\end{equation}
\vspace{-2mm}

The first term, $\mathbb{I}_{\text{attr}}(T, S)$, is a binary indicator that equals 1 if the subset $S$ covers the data attributes required by target insight $T$, and 0 otherwise.
The second term selects the most relevant insight within the subset $S$. 
It is jointly determined by the insight $I$'s statistical strength, $s_{\text{stat}}(I)$, and its content relevance to the target $T$, $s_{\text{cont}}(T, I)$.
We compute $s_{\text{cont}}(T, I)$ as the cosine similarity between the text embeddings of the task and insight description, generated by the text-embedding-3-small model~\cite{openai2024text}.
To restrict consideration to appropriate insight types, we set $s_{\text{cont}}(T, I) = -\infty$ when the insight type does not match the task.

For each target data insight, we select the top ten subsets with the highest matching scores to provide diverse candidates for composition. 
Following Luo~\etal~\cite{luo2018deepeye}, we employ a decision tree model to determine appropriate chart types (\eg bar chart and pie chart) for the selected subsets. 
These subsets and corresponding charts serve as basic building blocks for composition.
\begin{figure*}[!tb]
    \centering
    \includegraphics[width=\textwidth]{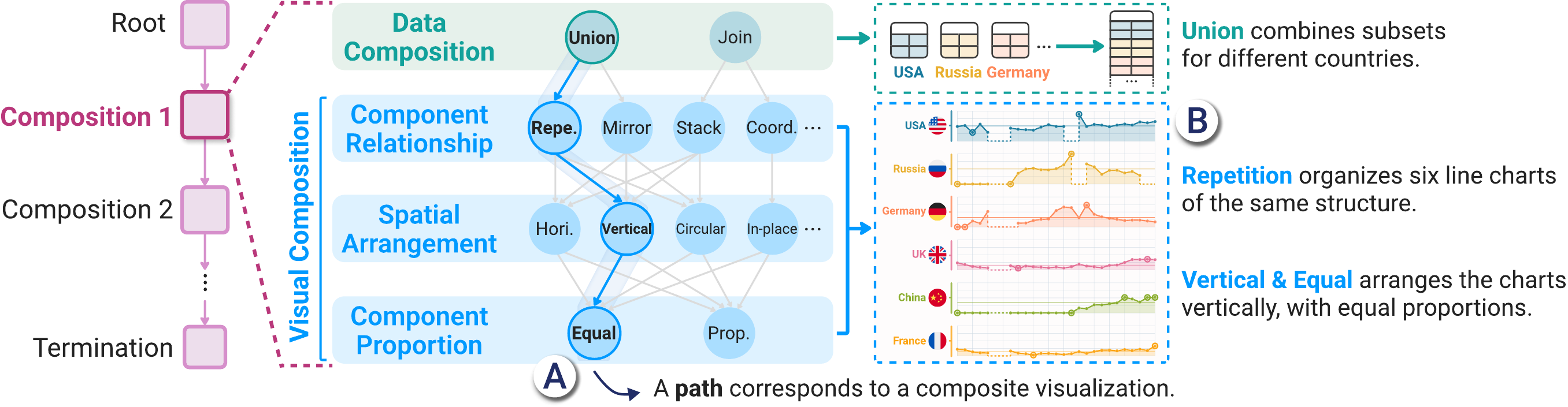}
    \caption{The composition graph models the data and visual compositions, as well as the feasible transitions between them.}
    \label{fig:formulation}
    \vspace{-4mm}
\end{figure*}
\subsection{MCGS-Based Visualization Composition}

In this section, we take the composite visualization design in Fig.~\ref{fig:teaser} as an example to illustrate the basic idea of our composition method.

\subsubsection{Problem Formulation}
Since visualization composition requires coordinated decisions over data compositions and visual compositions, we formulate it as a stepwise search problem over a composition graph (Fig.~\ref{fig:formulation}).
In this graph, each node represents either a data composition operation that specifies which data subsets are combined, or a visual composition operation that determines the component relationship and layout.
Each directed edge encodes a feasible transition between operations, reflecting compatibility constraints in the composition process. 
For example, after selecting the \textit{Repetition} component relationship, the subsequent spatial arrangement operation can be chosen from options such as \textit{Horizontal}, \textit{Vertical}, or \textit{Circular}, while options such as \textit{In-place} are not allowed.
Under this formulation, a path through the graph (Fig.~\ref{fig:formulation}A) corresponds to a composite visualization design (Fig.~\ref{fig:formulation}B).

\stitle{Data composition}.
To explicitly model how data relationships among components support coherent representations, we introduce two data composition operations: \textit{Union} and \textit{Join}.

\begin{figure}[!tb]
    \centering
    \includegraphics[width=0.488\textwidth]{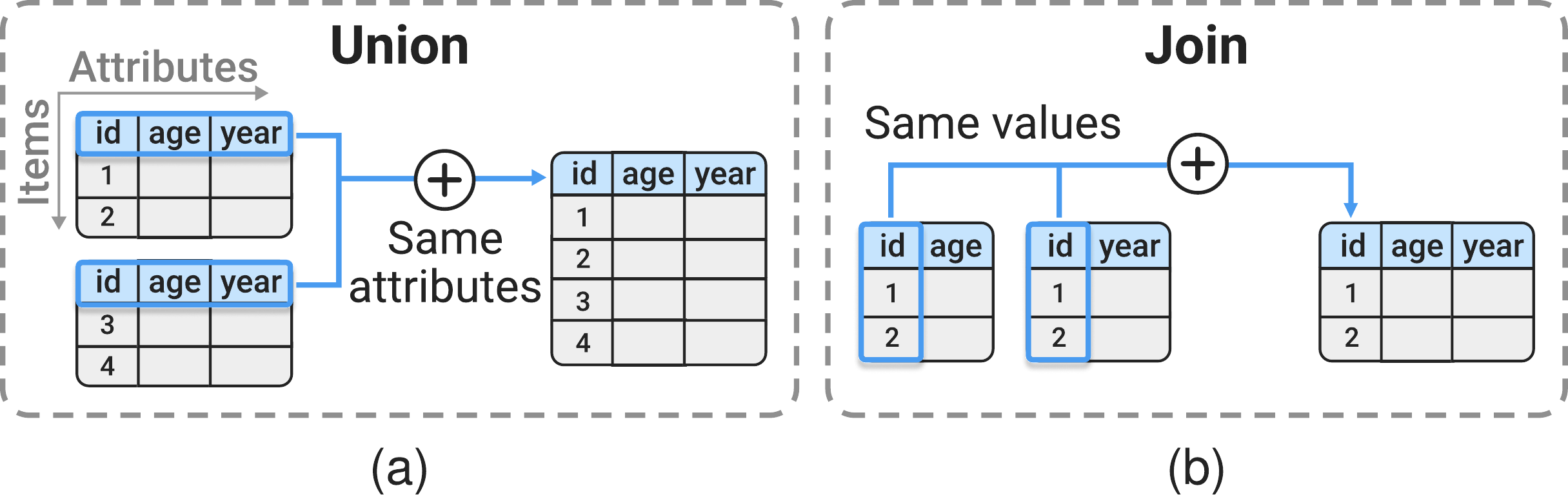}
    \caption{Two data composition operations: (a) Union, which combines data subsets sharing the same set of data attributes; (b) Join, which combines subsets sharing the same data attribute values.}
    \label{fig:data-composition}
    \vspace{-4mm}
\end{figure}

\begin{itemize}[nosep, wide=\parindent]
    \item \textit{Union} (Fig.~\ref{fig:data-composition}(a)) combines data subsets that share the same set of data attributes, and increases the number of data items by merging the rows of the subsets.
    For example, in Fig.~\ref{fig:teaser}B, unioning the subsets for different countries results in small multiples of bar charts.
    \item \textit{Join} (Fig.~\ref{fig:data-composition}(b)) combines subsets that share the same data attribute values, and increases data dimensions by merging attributes from different subsets.
    For example, in Fig.~\ref{fig:teaser}A\textasciitilde D, joining a group of subsets with the same values of the ``country'' attribute leads to the integration of the corresponding line charts, bar charts, and the alluvial diagram.
\end{itemize}

These two operations act as operators for subset combinations and build data relationships that enable visualization composition.
In the composition graph, each group of subsets that can be combined through either \textit{Union} or \textit{Join} corresponds to a data composition node.

\stitle{Visual composition}.
After combining the data subsets, we determine the component relationship, spatial arrangement, and component proportion according to the extended design space.
For example, the composition in Fig.~\ref{fig:teaser}B is specified as the tuple (\textit{Repetition}, \textit{Vertical}, \textit{Equal}), which includes six bar charts arranged in a vertical repetition, with equal space allocated to each chart.
Similarly, the composition in Fig.~\ref{fig:teaser}A\textasciitilde D is specified as (\textit{Stack}, \textit{Horizontal}, \textit{Proportional}).

\stitle{Operation transitions}.
To ensure that composite visualizations are structurally well-formed, we incorporate two compatibility constraints derived from our analysis of the composite visualization corpus to determine the feasible transitions between operations: \textit{data-visual operation compatibility} and \textit{visual representation compatibility}.

\begin{itemize}[nosep, wide=\parindent]
    \item \textit{Data-visual operation compatibility} specifies which data composition operations can support which component relationships. 
    Specifically, \textit{Union} supports \textit{Repetition} and \textit{Mirror} because these component relationships require data subsets with the same attributes. 
    \textit{Join} supports the remaining six component relationships, including \textit{Stack}, \textit{Co-axis}, \textit{Coordinate}, \textit{Annotation}, \textit{Large panel}, and \textit{Nesting}, because they combine subsets based on the same attribute values.
    
    \item \textit{Visual representation compatibility} specifies the feasible spatial arrangements and component proportions for each component relationship, as summarized in Fig.~\ref{fig:designspace}B. 
    For example, the \textit{Repetition} relationship supports arrangements such as \textit{Horizontal}, \textit{Vertical}, and \textit{Circular}, but not \textit{In-place}, since \textit{Repetition} requires non-overlapping juxtaposition, whereas \textit{In-place} involves spatial overlap.
    Similarly, the \textit{Co-axis} relationship supports only the \textit{Equal} proportion, since overlaid components share the same spatial region and coordinate system.
\end{itemize}

\subsubsection{MCGS Algorithm}
\label{subsubsec:algorithm}

As the search space grows exponentially with the number of composition operations, exhaustive search is computationally infeasible.
Prior systems have shown that Monte Carlo search is effective for exploring large search spaces in visualization creation~\cite{shi2021calliope, xie2024haichart}.
Motivated by these works, we employ MCGS~\cite{leurent2020monte} for visualization composition, because the composition process can be formulated as a graph-structured search problem with shared intermediate states.
MCGS can exploit this structure by sampling promising candidates, reusing search statistics across shared states, and balancing exploration and exploitation under a limited search budget.
Central to MCGS is a reward function that quantitatively evaluates the quality of composite visualization designs and guides the search toward more promising solutions.

\stitle{Algorithm overview}.
The algorithm iteratively proceeds through four phases: selection, expansion, simulation, and backpropagation.

\smash[b]{\underline{Selection}}.  
Starting from the root node, the algorithm traverses the composition graph by repeatedly selecting the node that best balances the exploitation of known, high-reward choices with the exploration of less-visited ones.

\smash[b]{\underline{Expansion}}.  
Once an unexplored node is reached, the algorithm identifies its unvisited subsequent nodes and randomly selects one to expand.
\looseness=-1

\smash[b]{\underline{Simulation}}.
To obtain a reward for the newly expanded node, this phase proceeds through four steps: \textit{path sampling}, \textit{layout optimization}, \textit{rendering}, and \textit{evaluation}.
In \textit{path sampling}, the algorithm randomly selects subsequent nodes to generate a path corresponding to a candidate composition, which is then mapped to a concrete layout.
In \textit{layout optimization}, we adopt the constrained packing formulation of Li~\etal~\cite{li2026chartgalaxy}, which maximizes the data-ink ratio while satisfying readability constraints, such as padding and alignment. 
We also apply consistency constraints, such as element ordering and axis configuration, to preserve semantic and structural coherence across components.
To efficiently solve this problem, we first generate an initial layout using a top-down greedy strategy, followed by local adjustments to satisfy the constraints and optimize the data-ink ratio.
The optimized layout is then \textit{rendered} using D3.js~\cite{bostock2011d3}, where colors, fonts, and chart styles are selected from our collected design examples to ensure visual harmony.
The rendered visualization is \textit{evaluated} by the reward function introduced below.

\smash[b]{\underline{Backpropagation}}.
The obtained reward for the expanded node is passed backward to the root, with each node along the path updating its reward to dynamically guide future search directions.

The four phases are repeated until the predefined iteration number is reached, after which the highest-reward candidate compositions are returned.
Further details of the algorithm are provided in Supp.~C, and the runtime analysis is provided in Supp.~D.

\stitle{Reward function}.
The reward function jointly optimizes task relevance ($R_{\text{task}}$), perceptual effectiveness ($R_{\text{eff}}$), and aesthetic coherence ($R_{\text{aes}}$):

\vspace{-2mm}
\begin{equation}
\text{Reward}=\alpha_1 {R_{\text{task}}}+\alpha_2 R_{\text{eff}}+\alpha_3 R_{\text{aes}},
\end{equation}
where $R_{\text{task}}$, $R_{\text{eff}}$, and $R_{\text{aes}}$ are normalized to the range $[0, 1]$, and the weighting parameters $\alpha_1=0.3$, $\alpha_2=0.5$, and $\alpha_3=0.2$ are determined through a grid search, which is detailed in Supp.~E.

\smash[b]{\underline{Task relevance}} measures how well a composite visualization design conveys the insights most relevant to the user's analysis task. 
Given an analysis task with $L$ target insights $\{T_1, T_2, \cdots, T_L\}$ and a composite visualization design comprising $N$ data subsets $\{S_1, S_2, \cdots, S_N\}$, {$R_{\text{task}}$} is computed based on the highest matching score for each target insight:

\vspace{-2mm}
\begin{equation}
    R_{\text{task}} =\textstyle \mathbb{I}_{\text{cov}} \cdot
    \sum_{j=1}^{L}w_j\max_{1\le i\le N} M(T_j, S_i).
\end{equation}

Here, $M(T_j, S_i)$ denotes the matching score between target insight $T_j$ and subset $S_i$, as described in Sec.~\ref{sec:selection}. 
The weight $w_j$ denotes the user-specified importance of target insight $T_j$, with a default value of $1/L$.
The binary indicator $\mathbb{I}_{\text{cov}}$ ensures insight coverage.
It equals 1 if the composite visualization design conveys all the target insights in the analysis task, and 0 otherwise.

\smash[b]{\underline{Perceptual effectiveness}} measures whether the layout of visual components facilitates efficient perception and interpretation~\cite{chen2020composition, wang2000guidelines, li2026NCP}.
Prior work by Zhou~\etal~\cite{zhou2024cluster} quantifies this aspect based on three layout-relevant Gestalt principles: \textit{proximity}, \textit{compactness}, and \textit{convexity}.
In particular, \textit{proximity} ($R_{\text{prox}}$) requires similar components (sharing the same data attributes or values) to be placed close together, \textit{compactness} ($R_{\text{comp}}$) ensures that components are arranged in a compact form, and \textit{convexity} ($R_{\text{conv}}$) indicates that components collectively form a convex shape.
We adopt the quantitative measures used by Zhou~\etal~\cite{zhou2024cluster}, corresponding to these three principles.

While these Gestalt-based measures capture perceptual grouping and spatial organization of components, they do not address how visual space should be proportionally allocated.
This limitation becomes particularly critical in composite visualizations, where multiple components must share limited display space, making space allocation a key factor that influences perceptual clarity, information retrieval efficiency, and analytical performance~\cite{chen2020composition}.
Generally, space allocation should reflect \textit{component importance}, which is jointly determined by the amount of information conveyed by the component (\ie the number of its data items) and its task priority (\ie the weight of its corresponding target insight)~\cite{cockburn2009review}.
Components with high importance should be allocated sufficient space to present detailed data insights, while others are assigned less space to provide supporting context~\cite{munzner2014visualization}.
To this end, we introduce a new measure, \textit{information balance} ($R_{\text{bal}}$), to quantify the alignment between each component's importance and its allocated space.
Following common practice~\cite{zhou2025hierarchical, yang2022diagnosing}, we compute this measure using KL-divergence~\cite{kullback1951information}:\looseness=-1
\vspace{-1mm}
\begin{equation}
    R_\text{bal} = \textstyle\exp\left(-\sum_{1\leq i\leq N} p_i\cdot\log \left(p_i / q_i\right) \right),
\end{equation}
where $p_i$ and $q_i$ denote the relative importance and the proportion of allocated space for the $i$-th component, respectively.
Both are normalized to sum to one (\ie $\sum_{i=1}^N p_i = \sum_{i=1}^N q_i = 1$), acting as discrete distributions for the KL-divergence.
The transformation $x \mapsto \exp (-x)$ is applied to normalize $R_{\text{bal}}$ to the range $[0, 1]$, following the practice of Zhou~\etal~\cite{zhou2024cluster}.

Perceptual effectiveness is calculated as a weighted sum of the aforementioned four measures:
\vspace{-1mm}
\begin{equation}
    R_{\text{eff}} = \beta_1 R_{\text{prox}} + \beta_2 R_{\text{comp}} + \beta_3 R_{\text{conv}} + \beta_4 R_{\text{bal}},
\end{equation}
where the weighting parameters $\beta_1$, $\beta_2$, $\beta_3$, and $\beta_4$ control the trade-off among the four measures. 
In our implementation, they are set to $0.2$, $0.3$, $0.3$, and $0.2$, respectively, through a grid search.
Further details about these measures and the grid search process are provided in Supp.~E.

\smash[b]{\underline{Aesthetic coherence}} assesses whether a composite visualization design maintains visual harmony and stylistic consistency across its components~\cite{harrison2015infographic, borkin2013makes}.
Since aesthetic coherence is subjective and hard to quantify, we employ the multimodal large language model (MLLM)-as-a-Judge paradigm~\cite{chen2024mllm, chen2024viseval, yang2024foundation} to approximate human judgments by leveraging its ability to reason over visual attributes (\eg color, typography).
Specifically, we use the Gemini-3-Flash model~\cite{gemini3} to assess three aspects of cross-component coherence: color-palette harmony, typography consistency, and uniformity of element styling.
The model outputs a rating on a scale ranging from 1 to 5, which is then normalized to $[0, 1]$.
Supp.~F provides the full prompt and empirically validates the alignment between MLLM-based judgments and human judgments.

\section{Evaluation}

We showcase the capabilities of \sys through a use case, and evaluate the reward function and composite visualizations via a user study.\looseness=-1

\subsection{Use Case}
This use case shows how \sys facilitates analytical communication and visual storytelling.
We implemented \sys as a web-based interface to support user interaction, as described in Supp.~G.
David, the publicity director of a college sports association, aims to design composite visualizations for an Olympic-themed poster for the campus sports culture week.
His goal is to present an engaging story about how Olympic participation and achievements have evolved over the past 100 years, highlighting patterns that resonate with the student audience.\looseness=-1

\begin{figure}[!tb]
    \centering
    \includegraphics[width=0.489\textwidth]{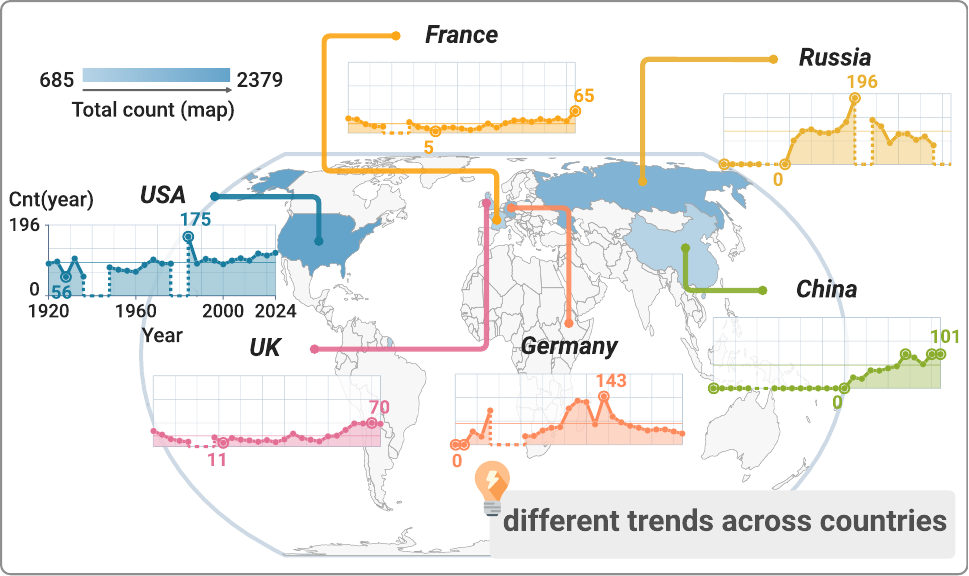}
    \caption{An alternative composite visualization design for the Summer Olympic Medals dataset: a world map overlaid with line chart annotations depicting medal trends across \textit{Countries}.}
    \label{fig:usecase}
    \vspace{-4mm}
\end{figure}

David reviews the data attributes and identifies several potential topics for the poster, such as country performance, gender differences, and sport distributions.
Among these topics, country-level analysis emerges as the most natural entry point, as it offers a global and competitive perspective of the Olympic Games.

\stitle{Investigating medal trends by country}.
To explore this topic, David first seeks to investigate the historical medal trends across countries.
He specifies the initial task by selecting the insight type \textit{trend} and data attributes \textit{Year} and \textit{Country}, resulting in the task: ``Show medal trends over \textit{Years} by \textit{Country}.'' 
Among the five candidate designs recommended by \sys, David agrees that the top two ranked designs are the most effective, as they best support cross-country comparison of medal trends over time.
The first design overlays six line charts representing different countries on a world map (Fig.~\ref{fig:usecase}), and the second design displays the same six line charts in a vertically aligned layout (Fig.~\ref{fig:teaser}A). 
Both designs effectively show the medal trends for different countries.
However, the second design provides clearer temporal alignment, making year-by-year comparisons more direct and interpretable. 
This alignment also makes key historical events more salient, such as the cancellation of the 1940 and 1944 Olympic Games during World War II and the Olympic boycotts by the USA (1980) and Russia (1984) during the Cold War. 
In contrast, the map-based layout spatially separates the time series, making such comparisons less immediate. 
David prefers the second design, as its aligned layout better supports year-by-year comparison and facilitates the identification of temporally aligned patterns and anomalies.
\looseness=-1

\stitle{Expanding the country-level analysis}.
To enrich the country-level narrative, David discusses with other members of the sports association and refines the task by adding ``Show the number values of each \textit{Medal Type} by \textit{Country}; and Compare \textit{Sex} differences for each \textit{Country}.''
\sys then recommends the top five ranked designs, including the design in Fig.~\ref{fig:teaser}A\textasciitilde C and the other four designs in Supp.~H.
In the first design, two additional sets of bar charts (Fig.~\ref{fig:teaser}B and C) reveal country-level differences in medal type and gender.
David observes that the USA and Russia maintain a high gold-medal ratio while leading in total medal counts, indicating their strong competitiveness. 
In terms of gender, most countries have won more medals from male athletes, whereas China stands out as the only country with more medals from female athletes.\looseness=-1

Continuing this iterative process, David further extends the task with: ``Show medal distributions by \textit{Sport} for each \textit{Country}.'' 
Among the recommended designs, he selects the one in Fig.~\ref{fig:teaser}(b), as it integrates multiple charts into a clear and coherent layout.
In this design, the added alluvial diagram (Fig.~\ref{fig:teaser}D) highlights each country's sporting strengths, such as the USA's dominance in \textit{Aquatics} and \textit{Athletics}, and Russia's prominence in \textit{Combat}.
Through these iterations, David develops a comprehensive design that effectively communicates a coherent and engaging country-level narrative for his poster.
This case demonstrates that \sys can help users with limited design knowledge to efficiently create composite visualizations for given tasks.

\subsection{User Study}
The study had two goals: 1) examining whether our reward function aligns more closely with human judgments than the state-of-the-art method, MLLM-as-a-Judge~\cite{chen2024mllm};
2) assessing whether users perceive the visualizations constructed by \sys as high-quality in terms of task relevance, perceptual effectiveness, and aesthetic coherence.
Our study was approved by our university’s IRB (No. THU-03-2025-1020).
Before the study began, participants were provided with an information sheet outlining the study plan and signed an informed consent form.

\subsubsection{User Study Design}
\stitle{Participants}.
We recruited 20 participants (15 males and 5 females) for our study. 
They were aged from 21 to 32 years old ($\textit{mean} = 24.4, \textit{SD} = 3.05$).
Their academic backgrounds included computer science (16) and information design (4), and their visualization design experience ranged from 1 to 10 years ($\textit{mean} = 3.33, \textit{SD} = 2.43$).
Each participant received \$30 compensation upon completion, independent of their performance.

\stitle{Datasets}.
We used three datasets: 1) Cars~\cite{cars}, 2) Spotify Songs~\cite{spotify}, and 3) Summer Olympic Medals~\cite{olympics}.
They have been widely used in prior works~\cite{wu2021multivision, wang2020datashot, huang2026tidynote} and cover different data types, including categorical, numerical, temporal, geometric, and relational data.
Details of these datasets are provided in Supp.~I.

\begin{figure*}[!b]
    \centering
    \vspace{-2mm}
    \includegraphics[width=0.99\textwidth]{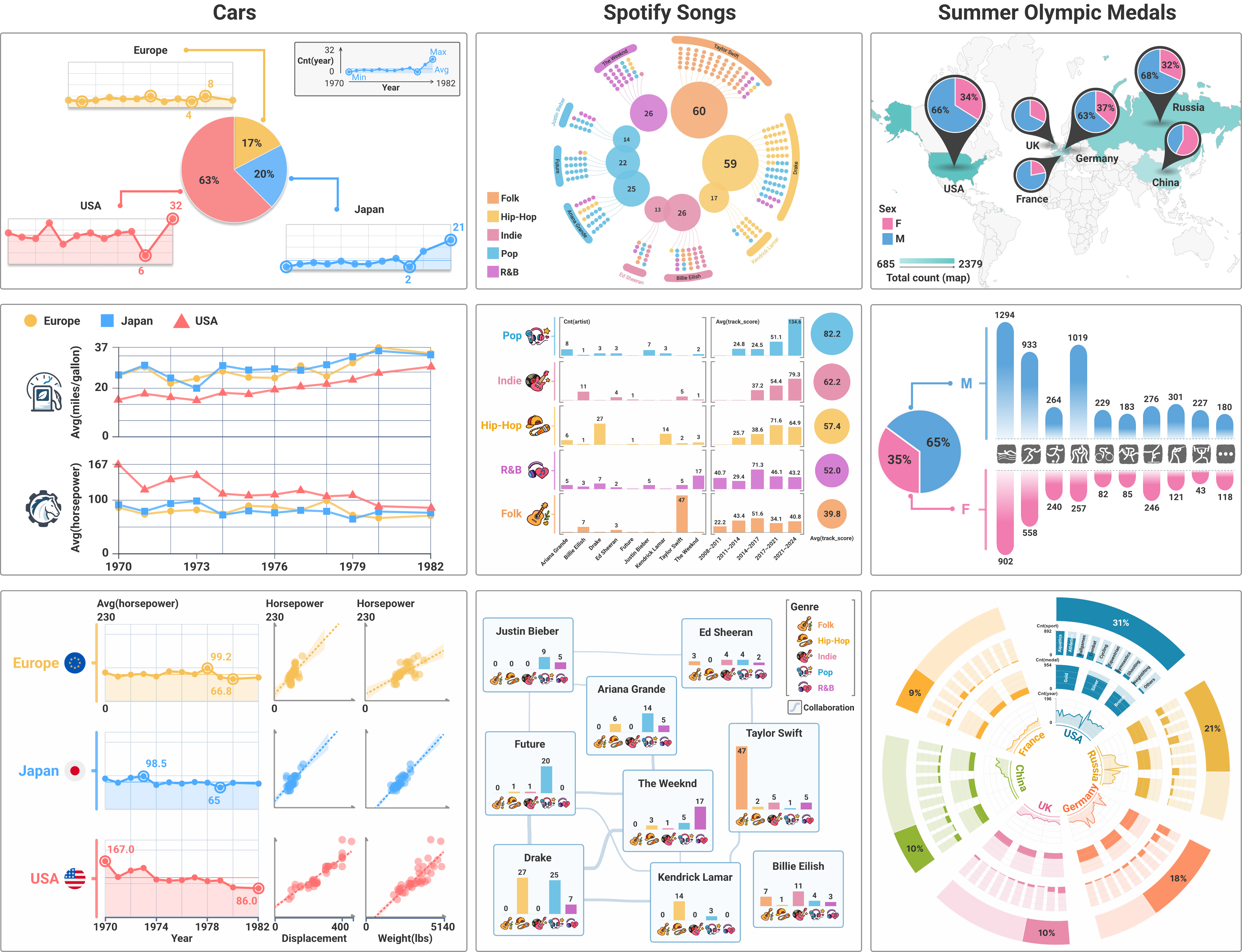}
    \caption{Composite visualization design examples used in the user study. From left to right, the examples correspond to the Cars, Spotify Songs, and Summer Olympic Medals datasets, respectively.}
    \label{fig:userstudyexample}
\end{figure*}

\stitle{Experiment design}.
We conducted two complementary experiments to examine both the alignment between different evaluation methods and human judgments (Experiment 1) and the perceived quality of the visualizations constructed by \sys (Experiment 2).

\smash[b]{\underline{Experiment 1}}.
To ensure the coverage of data types and analysis contexts, we designed two analysis tasks for each of the three datasets (six in total) and constructed composite visualization designs using \sys. 
Participants then judged these designs, and we compared their judgments with the scores from our reward function and MLLM-as-a-Judge.\looseness=-1

We collected human judgments via pairwise comparisons, which have been shown to provide more reliable preference signals than independent ratings~\cite{ouyang2022training, kiritchenko2017best}.
For each task, the participants compared the same 10 randomly sampled design pairs, resulting in $6(\textit{tasks})\times 10(\textit{pairs})=60$ comparisons.
For each pair, they selected the better design (or a tie) based on task relevance, perceptual clarity, and aesthetics. 
Responses were aggregated using majority voting~\cite{heer2010crowdsourcing}, where the majority choice was treated as the group consensus and vote differences as the voting margin.
\looseness=-1

We compared human judgments with our reward function and MLLM-as-a-Judge.
For the latter, we used Gemini-3-Flash~\cite{gemini3} and GPT-5.2~\cite{gpt5} due to their strong multimodal reasoning capabilities. 
Following prior works on MLLM-as-a-Judge~\cite{zheng2023judging, chen2024mllm}, we prompted the models to rate the visualization quality on a scale from 1 to 5, given the visualization, analysis task, and evaluation criteria. 
Each visualization design was assessed three times, and the scores were averaged to reduce stochasticity.
The prompt is provided in Supp.~I.
We quantified the alignment with human judgments using two complementary measures: 1) agreement rate with the human consensus, and 2) Spearman's rank correlation between score differences and human voting margins~\cite{luera2025mllm}.

\smash[b]{\underline{Experiment 2}}.
To avoid potential carryover effects from Experiment~1, we designed two new analysis tasks for each of the three datasets that are distinct from those used in Experiment 1 (six in total). 
To simulate real-world design scenarios, in which users select from a small set of candidate designs, the participants were presented with the top five ranked designs constructed by \sys for each task.
They selected the best design and rated its quality on a 7-point Likert scale across four dimensions: task relevance (``The design conveys the target data insights for the task''), readability (``The text, labels, and charts in the design are clear and easy to distinguish''), understandability (``I can easily and quickly understand the information and data insights presented in the design''), and aesthetics (``The overall design is visually pleasing''). 
Following prior works on visualization design~\cite{shi2021calliope, cabouat2024previs}, readability and understandability served as measures of perceptual effectiveness.
In total, we collected $20(\textit{participants})\times 6(\textit{tasks})\times 4(\textit{dimensions}) = 480$ ratings.\looseness=-1

The study lasted approximately 50 minutes, including a 10-minute training session and a 40-minute experiment session.
Some example designs are shown in Fig.~\ref{fig:userstudyexample}, with additional examples in Supp.~I.

\subsubsection{Result Analysis}
We analyze the results of the two experiments below.
Additional discussion informed by expert interviews is provided in Sec.~\ref{sec:discussion}.

\begin{figure}[!t]
    \centering
    \includegraphics[width=0.489\textwidth]{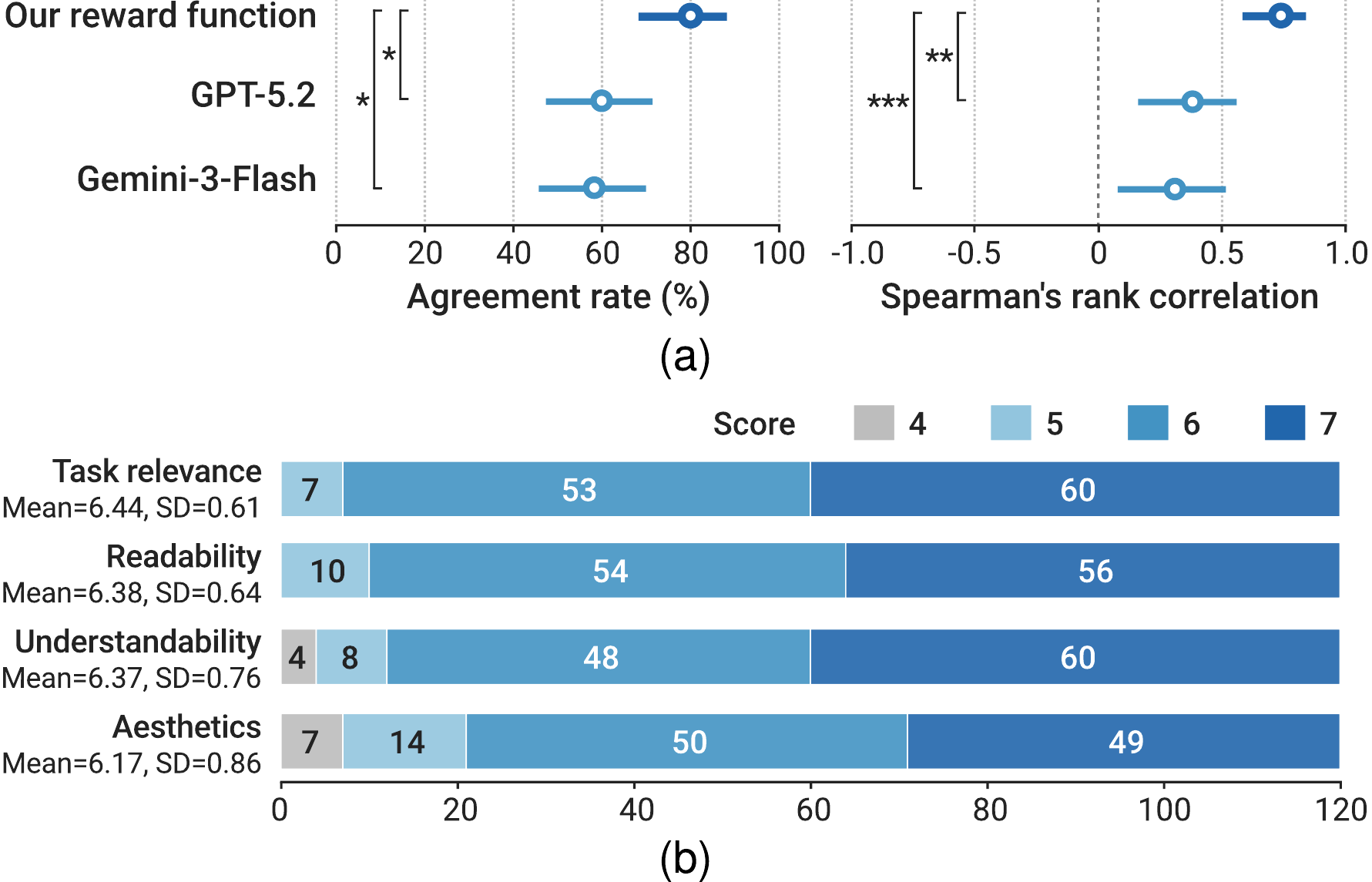}
    \caption{User study results: (a) the agreement rate and Spearman's rank correlation of our reward function and the MLLM-as-a-Judge baselines with human judgments; (b) user ratings for the selected designs across four dimensions. $*$ indicates $p < 0.05$, $**$ indicates $p < 0.01$, and $***$ indicates $p < 0.001$, 
    where $p$ is the Bonferroni-adjusted p-value.}
    \label{fig:us}
    \vspace{-4mm}
\end{figure}

\stitle{Alignment with human judgments}. 
Fig.~\ref{fig:us}(a) shows that our reward function aligns better with human judgments than MLLM-as-a-Judge.
It achieves an agreement rate of 80.0\%, outperforming GPT-5.2 (60.0\%) and Gemini-3-Flash (58.3\%).
McNemar's tests with the Bonferroni correction confirm that these differences are statistically significant ($\chi^2(1) = 6.05$ against GPT-5.2 and $\chi^2(1) = 6.86$ against Gemini-3-Flash, both adjusted $p < 0.05$).
A similar pattern is observed for Spearman's rank correlation: our reward function shows a stronger correlation with human judgments ($\rho = 0.739$) than GPT-5.2 ($\rho = 0.380$) and Gemini-3-Flash ($\rho = 0.309$).
Bootstrap tests (10,000 resamples) with the Bonferroni correction further confirm the significance of these differences (adjusted $p < 0.01$ \vs GPT-5.2; adjusted $p < 0.001$ \vs Gemini-3-Flash).
These results indicate that our reward function aligns more closely and consistently with human judgments.

\stitle{Perceived visualization quality}.
Fig.~\ref{fig:us}(b) summarizes the participant ratings.
The best designs selected by all the participants receive consistently high scores across the four dimensions: task relevance ($\textit{mean}=6.44, \textit{SD}=0.61$), readability ($\textit{mean}=6.38, \textit{SD}=0.64$), understandability ($\textit{mean}=6.37, \textit{SD}=0.76$), and aesthetics ($\textit{mean}=6.17, \textit{SD}=0.86$).
These results indicate that the constructed visualizations achieve a balanced performance across task relevance, perceptual effectiveness, and aesthetic coherence.
Among the four dimensions, aesthetics receives slightly lower scores and higher variance.
This may reflect the inherently subjective and context-dependent nature of aesthetic coherence, which is more difficult to model than task relevance and perceptual effectiveness.
\looseness=-1

\section{Expert Feedback and Discussion}
\label{sec:discussion}
After the user study, we conducted interviews with seven experts, including the four experts $E_1$--$E_4$ introduced in Sec.~\ref{sec:designgoals}, and three experts ($E_5$--$E_7$) who participated in the user study.
$E_5$ is a Ph.D. student in visual analytics, while $E_6$ and $E_7$ are Ph.D. students in information design, with 4.5, 4, and 5 years of visualization design experience, respectively.
The interviews focused on experts' overall experience with \sys and the usefulness of the constructed composite visualizations for analysis tasks.
Each interview lasted between 50 and 70 minutes.\looseness=-1

Overall, the experts reported positively on the usability of \sys.
They noted that it constructs high-quality compositions that clearly reveal data insights, maintain well-structured layouts, and exhibit strong visual coherence. 
Notably, the automatic chart selection and composition were highlighted as the major advantages.
These capabilities enable the experts to enhance both the efficiency and quality of the design process.
As a result, they can devote more attention to higher-level tasks such as data analysis and storytelling.
Furthermore, the experts emphasized that \sys offers a diverse range of candidate designs, providing a rich source of inspiration for accelerating ideation and prototyping.

Despite these benefits, the experts identified three key research opportunities to broaden the scope of our method further. 

\stitle{Combining with insight recommendation}.
Currently, \sys requires users to explicitly specify target data insights as part of the analysis tasks.
However, all experts commented that this requirement may limit its usability in exploratory settings, where users are unfamiliar with the dataset or lack well-defined analytical goals.
To better support such scenarios, the tool should automatically identify potentially interesting insights and further organize them into meaningful analysis tasks.
A straightforward solution is to integrate with existing insight recommendation methods~\cite{ding2019quickinsights, aodeng2025inreactable, zhao2026proactive}.
However, these methods often produce fragmented insights that lack the analytical connections necessary for task formulation.
This limitation highlights the need to reason about the relationships among candidate insights and structure them into well-formed analysis tasks.

\stitle{Incremental composition}.
From a practical perspective, most experts highlighted the need for user-controllable composition.
For example, $E_3$ suggested: ``In some cases, I arrive at an intermediate design that I would like to preserve. 
It would be useful if \sys can fix this partial result and continue exploring alternatives for the remaining components.''
Similarly, $E_7$ expressed the need to support incremental design: ``In practice, I tend to design the core components first and then iteratively refine the rest. 
Supporting such incremental composition would make \sys more applicable to real-world use.''
These requirements suggest a promising research direction for developing user-guided search mechanisms that dynamically adjust the search space in response to user input.
This can be achieved by mapping user requirements (\eg fixing specific components) as search constraints, or by integrating user feedback into the reward function to guide the search direction.\looseness=-1

\stitle{Continually enriching the corpus}.
All experts noted that \sys can construct diverse composite visualizations.
This highlights the value of corpus diversity, which provides the basis for our composition method.
While our corpus draws on three complementary sources and includes a quality control process, it remains a snapshot of existing public composite visualization designs.
As visualization design continues to evolve, new composition strategies may emerge, including new component relationships, spatial arrangements, and component proportions.
This highlights the need to continually enrich the corpus.
As new composition strategies are identified, the design space and composition graph can be updated accordingly, preventing \sys from being limited to the strategies observed in the current corpus.

\section{Conclusion}

We introduce \sys, a task-aware visualization composition method that formulates the design process as a stepwise graph search problem.
Guided by an extended design space derived from analyzing \numDataset composite visualization designs, our method employs the MCGS algorithm to efficiently search for composition alternatives.
The search is guided by a reward function that balances task relevance, perceptual effectiveness, and aesthetic coherence.
We present a use case to illustrate its ability to construct composite visualizations that effectively support analytical communication and visual storytelling.
We further conduct a user study to demonstrate that the constructed designs align well with human judgments of quality.
These findings demonstrate the feasibility of modeling visualization composition as a structured search process grounded in visualization design principles.


\acknowledgments{
The authors would like to thank Zhen Li, Lanxi Xiao, and Duan Li for their valuable contributions to the discussions.
}

\bibliographystyle{abbrv-doi-hyperref}

\bibliography{reference}

@inproceedings{javed2012exploring,
  title={Exploring the design space of composite visualization},
  author={Javed, Waqas and Elmqvist, Niklas},
  booktitle={Proceedings of the IEEE Pacific Visualization Symposium},
  pages={1--8},
  year={2012},
  doi={10.1109/PacificVis.2012.6183556},
}

@article{deng2023revisiting,
  title={Revisiting the design patterns of composite visualizations},
  author={Deng, Dazhen and Cui, Weiwei and Meng, Xiyu and Xu, Mengye and Liao, Yu and Zhang, Haidong and Wu, Yingcai},
  journal={IEEE Transactions on Visualization and Computer Graphics},
  volume={29},
  number={12},
  pages={5406--5421},
  year={2023},
  doi={10.1109/TVCG.2022.3213565},
}

@article{zhu2024compositing,
  title={{CompositingVis}: Exploring interactions for creating composite visualizations in immersive environments},
  author={Zhu, Qian and Lu, Tao and Guo, Shunan and Ma, Xiaojuan and Yang, Yalong},
  journal={IEEE Transactions on Visualization and Computer Graphics},
  year={2025},
  volume={31},
  number={01},
  pages={591--601},
  doi={10.1109/TVCG.2024.3456210},
}

@inproceedings{luo2018deepeye,
  title={{DeepEye}: Towards Automatic Data Visualization},
  booktitle={Proceedings of the IEEE International Conference on Data Engineering},
  author={Luo, Yuyu and Qin, Xuedi and Tang, Nan and Li, Guoliang},
  year={2018},
  pages={101--112},
  doi={10.1109/ICDE.2018.00019},
}

@article{wang2020datashot,
  title={{DataShot}: Automatic Generation of Fact Sheets from Tabular Data},
  author={Wang, Yun and Sun, Zhida and Zhang, Haidong and Cui, Weiwei and Xu, Ke and Ma, Xiaojuan and Zhang, Dongmei},
  year={2020},
  journal={IEEE Transactions on Visualization and Computer Graphics},
  volume={26},
  number={1},
  pages={895--905},
  doi={10.1109/TVCG.2019.2934398},
}

@article{shi2021calliope,
  title={Calliope: Automatic Visual Data Story Generation from a Spreadsheet},
  author={Shi, Danqing and Xu, Xinyue and Sun, Fuling and Shi, Yang and Cao, Nan},
  year={2021},
  journal={IEEE Transactions on Visualization and Computer Graphics},
  volume={27},
  number={2},
  pages={453--463},
  doi={10.1109/tvcg.2020.3030403},
}

@article{wu2021multivision,
  author={Wu, Aoyu and Wang, Yun and Zhou, Mengyu and He, Xinyi and Zhang, Haidong and Qu, Huamin and Zhang, Dongmei},
  journal={IEEE Transactions on Visualization and Computer Graphics},
  title={{MultiVision}: Designing Analytical Dashboards with Deep Learning Based Recommendation},
  year={2022},
  volume={28},
  number={1},
  pages={162--172},
  doi={10.1109/TVCG.2021.3114826},
}

@article{lange2025aardvark,
  title={{Aardvark}: Composite visualizations of trees, time-series, and images},
  author={Lange, Devin and Judson-Torres, Robert and Zangle, Thomas A and Lex, Alexander},
  journal={IEEE Transactions on Visualization and Computer Graphics},
  volume={31},
  number={1},
  pages={1290--1300},
  year={2025},
  doi={10.1109/TVCG.2024.3456193},
}

@article{chen2021interactive,
  title={Interactive Graph Construction for Graph-Based Semi-Supervised Learning},
  author={Chen, Changjian and Wang, Zhaowei and Wu, Jing and Wang, Xiting and Guo, Lan-Zhe and Li, Yu-Feng and Liu, Shixia},
  journal={IEEE Transactions on Visualization and Computer Graphics},
  year={2021},
  volume={27},
  number={9},
  pages={3701--3716},
  doi={10.1109/TVCG.2021.3084694},
}

@article{bostock2011d3,
  author={Bostock, Michael and Ogievetsky, Vadim and Heer, Jeffrey},
  title={D3 Data-Driven Documents},
  year={2011},
  volume={17},
  number={12},
  journal={IEEE Transactions on Visualization and Computer Graphics},
  pages={2301--2309},
  doi={10.1109/TVCG.2011.185},
}

@inproceedings{ding2019quickinsights,
  title={{QuickInsights}: Quick and Automatic Discovery of Insights from Multi-Dimensional Data},
  booktitle={Proceedings of the International Conference on Management of Data},
  author={Ding, Rui and Han, Shi and Xu, Yong and Zhang, Haidong and Zhang, Dongmei},
  year={2019},
  pages={317--332},
  doi={10.1145/3299869.3314037},
}

@inproceedings{wang2000guidelines,
  author={Wang Baldonado, Michelle Q. and Woodruff, Allison and Kuchinsky, Allan},
  title={Guidelines for using multiple views in information visualization},
  year={2000},
  booktitle={Proceedings of the Working Conference on Advanced Visual Interfaces},
  pages={110--119},
  doi={10.1145/345513.345271},
}

@article{meyer2009mizbee,
  author={Meyer, Miriah and Munzner, Tamara and Pfister, Hanspeter},
  journal={IEEE Transactions on Visualization and Computer Graphics},
  title={{MizBee}: A Multiscale Synteny Browser},
  year={2009},
  volume={15},
  number={6},
  pages={897--904},
  doi={10.1109/TVCG.2009.167},
}

@article{yang2022diagnosing,
  author={Yang, Weikai and Ye, Xi and Zhang, Xingxing and Xiao, Lanxi and Xia, Jiazhi and Wang, Zhongyuan and Zhu, Jun and Pfister, Hanspeter and Liu, Shixia},
  journal={IEEE Transactions on Visualization and Computer Graphics},
  title={Diagnosing Ensemble Few-Shot Classifiers},
  year={2022},
  volume={28},
  number={9},
  pages={3292--3306},
  doi={10.1109/TVCG.2022.3182488},
}

@article{zhou2024cluster,
  title={Cluster-Aware Grid Layout},
  author={Zhou, Yuxing and Yang, Weikai and Chen, Jiashu and Chen, Changjian and Shen, Zhiyang and Luo, Xiaonan and Yu, Lingyun and Liu, Shixia},
  year={2024},
  journal={IEEE Transactions on Visualization and Computer Graphics},
  volume={30},
  number={1},
  pages={240--250},
  doi={10.1109/TVCG.2023.3326934},
}

@inproceedings{chen2024mllm,
  title={{MLLM-as-a-Judge}: Assessing multimodal {LLM-as-a-Judge} with vision-language benchmark},
  author={Chen, Dongping and Chen, Ruoxi and Zhang, Shilin and Wang, Yaochen and Liu, Yinuo and Zhou, Huichi and Zhang, Qihui and Wan, Yao and Zhou, Pan and Sun, Lichao},
  booktitle={Proceedings of the International Conference on Machine Learning},
  year={2024},
  pages={6562--6595},
}

@inproceedings{li2026chartgalaxy,
  title={{ChartGalaxy}: A Dataset for Infographic Chart Understanding and Generation},
  author={Li, Zhen and Li, Duan and Guo, Yukai and Guo, Xinyuan and Li, Bowen and Xiao, Lanxi and Qiao, Shenyu and Chen, Jiashu and Wu, Zijian and Zhang, Hui and Shu, Xinhuan and Liu, Shixia},
  booktitle={Proceedings of the International Conference on Learning Representations},
  year={2026},
}

@article{lex2014upset,
  title={{UpSet}: Visualization of Intersecting Sets},
  author={Lex, Alexander and Gehlenborg, Nils and Strobelt, Hendrik and Vuillemot, Romain and Pfister, Hanspeter},
  year={2014},
  journal={IEEE Transactions on Visualization and Computer Graphics},
  volume={20},
  number={12},
  pages={1983--1992},
  doi={10.1109/TVCG.2014.2346248},
}

@article{kullback1951information,
  author={Kullback, S. and Leibler, R. A.},
  title={On information and sufficiency},
  journal={Annals of Mathematical Statistics},
  year={1951},
  volume={22},
  number={1},
  pages={79--86},
  doi={10.1214/aoms/1177729694},
}

@article{chen2024viseval,
  author={Chen, Nan and Zhang, Yuge and Xu, Jiahang and Ren, Kan and Yang, Yuqing},
  journal={IEEE Transactions on Visualization and Computer Graphics},
  title={{VisEval}: A Benchmark for Data Visualization in the Era of Large Language Models},
  year={2025},
  volume={31},
  number={1},
  pages={1301--1311},
  doi={10.1109/TVCG.2024.3456320},
}

@inproceedings{leurent2020monte,
  title={{Monte-Carlo} Graph Search: the Value of Merging Similar States},
  author={Leurent, Edouard and Maillard, Odalric-Ambrym},
  booktitle={Proceedings of the Asian Conference on Machine Learning},
  pages={577--592},
  year={2020},
}

@article{liu2017towards,
  title={Towards Better Analysis of Deep Convolutional Neural Networks}, 
  author={Liu, Mengchen and Shi, Jiaxin and Li, Zhen and Li, Chongxuan and Zhu, Jun and Liu, Shixia},
  journal={IEEE Transactions on Visualization and Computer Graphics}, 
  year={2017},
  volume={23},
  number={1},
  pages={91--100},
  doi={10.1109/TVCG.2016.2598831}
}

@article{cui2011textflow,
  title={{TextFlow}: Towards Better Understanding of Evolving Topics in Text}, 
  author={Cui, Weiwei and Liu, Shixia and Tan, Li and Shi, Conglei and Song, Yangqiu and Gao, Zekai and Qu, Huamin and Tong, Xin},
  journal={IEEE Transactions on Visualization and Computer Graphics}, 
  year={2011},
  volume={17},
  number={12},
  pages={2412--2421},
  doi={10.1109/TVCG.2011.239}
}

@article{shi2026piccl,
  title={{PiCCL}: Data-Driven Composition of Bespoke Pictorial Charts}, 
  author={Shi, Haoyan and Wang, Yunhai and Chen, Junhao and Wang, Chenglong and Lee, Bongshin},
  journal={IEEE Transactions on Visualization and Computer Graphics}, 
  year={2026},
  volume={32},
  number={1},
  pages={714--724},
  doi={10.1109/TVCG.2025.3634264}
}

@article{zhao2026proactive,
  title={{ProactiveVA}: Proactive Visual Analytics with {LLM}-Based {UI} Agent}, 
  author={Zhao, Yuheng and Shu, Xueli and Fan, Liwen and Gao, Lin and Zhang, Yu and Chen, Siming},
  journal={IEEE Transactions on Visualization and Computer Graphics}, 
  year={2026},
  volume={32},
  number={1},
  pages={451--461},
  doi={10.1109/TVCG.2025.3642628}
}

@article{yang2025dashboard,
  title={Dashboard Vision: Using Eye-Tracking to Understand and Predict Dashboard Viewing Behaviors}, 
  author={Yang, Manling and Hou, Yihan and Li, Ling and Chang, Remco and Zeng, Wei},
  journal={IEEE Transactions on Visualization and Computer Graphics}, 
  year={2025},
  volume={31},
  number={10},
  pages={6930--6945},
  doi={10.1109/TVCG.2025.3532497}
}

@inproceedings{harrison2015infographic,
  title={Infographic aesthetics: Designing for the first impression},
  author={Harrison, Lane and Reinecke, Katharina and Chang, Remco},
  booktitle={Proceedings of the CHI Conference on Human Factors in Computing Systems},
  pages={1187--1190},
  year={2015},
  doi={10.1145/2702123.2702545},
}

@inproceedings{lu2020exploring,
  title={Exploring visual information flows in infographics},
  author={Lu, Min and Wang, Chufeng and Lanir, Joel and Zhao, Nanxuan and Pfister, Hanspeter and Cohen-Or, Daniel and Huang, Hui},
  booktitle={Proceedings of the CHI Conference on Human Factors in Computing Systems},
  pages={1--12},
  year={2020},
  doi={10.1145/3313831.3376263},
}

@inproceedings{zheng2023judging,
  author={Zheng, Lianmin and Chiang, Wei-Lin and Sheng, Ying and Zhuang, Siyuan and Wu, Zhanghao and Zhuang, Yonghao and Lin, Zi and Li, Zhuohan and Li, Dacheng and Xing, Eric P. and Zhang, Hao and Gonzalez, Joseph E. and Stoica, Ion},
  title={Judging {LLM-as-a-Judge} with {MT-Bench} and {Chatbot Arena}},
  year={2023},
  booktitle={Proceedings of the International Conference on Neural Information Processing Systems},
  pages={46595--46623},
}

@inproceedings{ying2024vaid,
  title={{VAID}: Indexing View Designs in Visual Analytics System},
  booktitle={Proceedings of the CHI Conference on Human Factors in Computing Systems},
  author={Ying, Lu and Wu, Aoyu and Li, Haotian and Deng, Zikun and Lan, Ji and Wu, Jiang and Wang, Yong and Qu, Huamin and Deng, Dazhen and Wu, Yingcai},
  year={2024},
  pages={1--15},
  doi={10.1145/3613904.3642237},
}

@article{henry2007nodetrix,
  title={{NodeTrix}: a hybrid visualization of social networks},
  author={Henry, Nathalie and Fekete, Jean-Daniel and McGuffin, Michael J},
  journal={IEEE Transactions on Visualization and Computer Graphics},
  volume={13},
  number={6},
  pages={1302--1309},
  year={2007},
  doi={10.1109/TVCG.2007.70582},
}

@article{chen2020composition,
  title={Composition and configuration patterns in multiple-view visualizations},
  author={Chen, Xi and Zeng, Wei and Lin, Yanna and Ai-Maneea, Hayder Mahdi and Roberts, Jonathan and Chang, Remco},
  journal={IEEE Transactions on Visualization and Computer Graphics},
  volume={27},
  number={2},
  pages={1514--1524},
  year={2021},
  doi={10.1109/TVCG.2020.3030338},
}

@article{gleicher2011visual,
  title={Visual comparison for information visualization},
  author={Gleicher, Michael and Albers, Danielle and Walker, Rick and Jusufi, Ilir and Hansen, Charles D and Roberts, Jonathan C},
  journal={Information Visualization},
  volume={10},
  number={4},
  pages={289--309},
  year={2011},
  doi={10.1177/1473871611416549},
}

@article{collins2007vislink,
  title={{VisLink}: Revealing relationships amongst visualizations},
  author={Collins, Christopher and Carpendale, Sheelagh},
  journal={IEEE Transactions on Visualization and Computer Graphics},
  volume={13},
  number={6},
  pages={1192--1199},
  year={2007},
  doi={10.1109/TVCG.2007.70521},
}

@inproceedings{heer2010crowdsourcing,
  title={Crowdsourcing graphical perception: using mechanical turk to assess visualization design},
  author={Heer, Jeffrey and Bostock, Michael},
  booktitle={Proceedings of the CHI Conference on Human Factors in Computing Systems},
  pages={203--212},
  year={2010},
  doi={10.1145/1753326.1753357},
}

@article{zhou2025hierarchical,
  author={Zhou, Yuxing and Chen, Changjian and Shen, Zhiyang and Zhu, Jiangning and Chen, Jiashu and Yang, Weikai and Liu, Shixia},
  journal={IEEE Transactions on Visualization and Computer Graphics},
  title={Hierarchical Fuzzy-Cluster-Aware Grid Layout for Large-Scale Data},
  year={2025},
  volume={31},
  number={10},
  pages={8200--8213},
  doi={10.1109/TVCG.2025.3566558},
}

@article{cockburn2009review,
  author={Cockburn, Andy and Karlson, Amy and Bederson, Benjamin B.},
  title={A review of overview+detail, zooming, and focus+context interfaces},
  year={2009},
  volume={41},
  number={1},
  journal={ACM Computing Surveys},
  pages={1--31},
  doi={10.1145/1456650.1456652},
}

@inproceedings{aodeng2025inreactable,
  author={Aodeng, Gerile and Li, Guozheng and Feng, Yunshan and Chen, Qiyang and Zhang, Yu and Liu, Chi Harold},
  title={{InReAcTable}: {LLM}-powered Interactive Visual Data Story Construction from Tabular Data},
  year={2025},
  booktitle={Proceedings of the Annual ACM Symposium on User Interface Software and Technology},
  pages={1--16},
  doi={10.1145/3746059.3747719},
}

@article{wang2018visualization,
  title={Visualization and visual analysis of ensemble data: A survey},
  author={Wang, Junpeng and Hazarika, Subhashis and Li, Cheng and Shen, Han-Wei},
  journal={IEEE Transactions on Visualization and Computer Graphics},
  volume={25},
  number={9},
  pages={2853--2872},
  year={2019},
  doi={10.1109/TVCG.2018.2853721},
}

@article{cabouat2024previs,
  title={{PREVis}: Perceived readability evaluation for visualizations},
  author={Cabouat, Anne-Flore and He, Tingying and Isenberg, Petra and Isenberg, Tobias},
  journal={IEEE Transactions on Visualization and Computer Graphics},
  volume={31},
  number={1},
  pages={1083--1093},
  year={2025},
  doi={10.1109/TVCG.2024.3456318},
}

@article{borkin2013makes,
  title={What makes a visualization memorable?},
  author={Borkin, Michelle A and Vo, Azalea A and Bylinskii, Zoya and Isola, Phillip and Sunkavalli, Shashank and Oliva, Aude and Pfister, Hanspeter},
  journal={IEEE Transactions on Visualization and Computer Graphics},
  volume={19},
  number={12},
  pages={2306--2315},
  year={2013},
  doi={10.1109/TVCG.2013.234},
}

@article{solen2026design,
  author={Solen, Mara and Oddo, Matt and Munzner, Tamara},
  journal={IEEE Transactions on Visualization and Computer Graphics},
  title={A Design Space for Multiscale Visualization},
  year={2026},
  volume={32},
  number={1},
  pages={1372--1382},
  doi={10.1109/TVCG.2025.3634790},
}

@article{yang2024foundation,
  title={Foundation models meet visualizations: Challenges and opportunities},
  author={Yang, Weikai and Liu, Mengchen and Wang, Zheng and Liu, Shixia},
  journal={Computational Visual Media},
  volume={10},
  number={3},
  pages={399--424},
  year={2024},
  doi={10.1007/s41095-023-0393-x},
}

@inproceedings{yao2025designing,
  title={Designing Visualization Widgets for Tangible Data Exploration: A Systematic Review},
  author={Yao, Haonan and Yu, Lingyun and Yao, Lijie},
  booktitle={Proceedings of the IEEE Visualization and Visual Analytics},
  year={2025},
  pages={261--265},
  doi={10.1109/VIS60296.2025.00058},
}

@inproceedings{ouyang2022training,
  author={Ouyang, Long and Wu, Jeff and Jiang, Xu and Almeida, Diogo and Wainwright, Carroll L. and Mishkin, Pamela and Zhang, Chong and Agarwal, Sandhini and Slama, Katarina and Ray, Alex and Schulman, John and Hilton, Jacob and Kelton, Fraser and Miller, Luke and Simens, Maddie and Askell, Amanda and Welinder, Peter and Christiano, Paul and Leike, Jan and Lowe, Ryan},
  title={Training language models to follow instructions with human feedback},
  year={2022},
  booktitle={Proceedings of the International Conference on Neural Information Processing Systems},
  pages={27730--27744},
}

@inproceedings{kiritchenko2017best,
  title={Best-worst scaling more reliable than rating scales: A case study on sentiment intensity annotation},
  author={Kiritchenko, Svetlana and Mohammad, Saif},
  booktitle={Proceedings of the Annual Meeting of the Association for Computational Linguistics},
  pages={465--470},
  year={2017},
  doi={10.18653/v1/P17-2074},
}

@article{li2026NCP,
  title={{NCP}: Neighborhood-Preserving Non-Uniform Circle Packing Method for Visualization},
  journal={Computational Visual Media},
  publisher={Springer Science and Business Media LLC},
  author={Duan Li and Jun Yuan and Xinyuan Guo and Xiting Wang and Yang Liu and Weikai Yang and Shixia Liu},
  year={2026},
  note={to appear},
}

@article{luera2025mllm,
  title={{MLLM} as a {UI} {Judge}: Benchmarking multimodal {LLMs} for predicting human perception of user interfaces},
  author={Luera, Reuben A. and Rossi, Ryan and Dernoncourt, Franck and Basu, Samyadeep and Kim, Sungchul and Mukherjee, Subhojyoti and Mathur, Puneet and Zhang, Ruiyi and Kil, Jihyung and Lipka, Nedim and Yoon, Seunghyun and Gu, Jiuxiang and Wang, Zichao and Bearfield, Cindy Xiong and Kveton, Branislav},
  journal={arXiv preprint arXiv:2510.08783},
  year={2025},
  doi={10.48550/arXiv.2510.08783},
}

@article{huang2026tidynote,
  title={Tidynote: Always-Clear Notebook Authoring},
  author={Huang, Ruanqianqian and Hempel, Brian and Cao, Yining and Hollan, James D and Xia, Haijun and Lerner, Sorin},
  journal={arXiv preprint arXiv:2602.23490},
  year={2026},
  doi={10.48550/arXiv.2602.23490},
}

@book{kirk2016data,
  author={Kirk, Andy},
  title={Data Visualisation: A Handbook for Data Driven Design},
  year={2016},
  isbn={1473912148},
  publisher={Sage Publications Ltd.},
}

@book{munzner2014visualization,
  title={Visualization Analysis and Design},
  author={Munzner, Tamara},
  year={2014},
  publisher={CRC press},
  doi={10.1145/3721241.3733989},
}

@book{playfair1801statistical,
  title={The statistical breviary; shewing the resources of every state and kingdom in Europe},
  author={Playfair, William},
  year={1801},
  publisher={John Wallis},
}

@misc{pinterest,
  title={{Pinterest}},
  author={{Pinterest}},
  year={2026},
  howpublished={https://www.pinterest.com/},
  note={Last accessed 2026-08-04},
}

@misc{visualcapitalist,
  title={{Visual Capitalist}},
  author={{Visual Capitalist}},
  year={2026},
  howpublished={https://www.visualcapitalist.com/},
  note={Last accessed 2026-08-04},
}

@misc{spotify,
  title={{Most Streamed Spotify Songs 2024}},
  author={Nidula Elgiriyewithana},
  year={2024},
  howpublished={https://www.kaggle.com/datasets/nelgiriyewithana/most-streamed-spotify-songs-2024/},
  note={Last accessed 2026-08-04},
}

@misc{olympics,
  title={{Summer Olympics Medals (1896-2024)}},
  author={Stefany De Oliveira},
  year={2024},
  howpublished={https://www.kaggle.\allowbreak com/\allowbreak datasets/\allowbreak stefanydeoliveira/\allowbreak summer-olympics-medals-1896-2024/},
  note={Last accessed 2026-08-04},
}

@misc{cars,
  title={{Vega Datasets}},
  author={Vega},
  year={2018},
  howpublished={https://github.com/vega/vega-datasets/},
  note={Last accessed 2026-08-04},
}

@misc{gemini3,
  title={{Gemini 3 Flash: frontier intelligence built for speed}},
  author={{Google}},
  year={2026},
  howpublished={https://blog.google/products-and-platforms/products/gemini/gemini-3-flash/},
  note={Last accessed 2026-08-04},
}

@misc{gpt5,
  title={{Introducing GPT-5.2}},
  author={{OpenAI}},
  year={2026},
  howpublished={https://openai.com/index/introducing-gpt-5-2/},
  note={Last accessed 2026-08-04},
}

@misc{openai2024text,
  author={{OpenAI}},
  title={{Text-embedding-3-small}},
  year={2024},
  howpublished={https://platform.\allowbreak openai.\allowbreak com/\allowbreak docs/\allowbreak models/\allowbreak text-embedding-3-small/},
  note={Last accessed 2026-08-04},
}

@article{munzner2009nested,
  title={A nested model for visualization design and validation},
  author={Munzner, Tamara},
  journal={IEEE Transactions on Visualization and Computer Graphics},
  volume={15},
  number={6},
  pages={921--928},
  year={2009},
  doi={10.1109/TVCG.2009.111},
}

@article{brehmer2013multi,
  title={A multi-level typology of abstract visualization tasks},
  author={Brehmer, Matthew and Munzner, Tamara},
  journal={IEEE Transactions on Visualization and Computer Graphics},
  volume={19},
  number={12},
  pages={2376--2385},
  year={2013},
  doi={10.1109/TVCG.2013.124},
}

@article{xie2024haichart,
  title={{HAIChart}: Human and {AI} Paired Visualization System},
  author={Xie, Yupeng and Luo, Yuyu and Li, Guoliang and Tang, Nan},
  journal={Proceedings of the VLDB Endowment},
  volume={17},
  number={11},
  pages={3178--3191},
  year={2024},
  doi={10.14778/3681954.3681992},
}

\end{document}